\documentclass[draftclsnofoot, onecolumn]{IEEEtran}
\usepackage{graphicx}
\usepackage{amsmath}
\usepackage{amssymb}
\usepackage{graphics}
\usepackage{latexsym}
\usepackage[dvips]{epsfig}
\usepackage{cite}
\usepackage{subfigure}
\usepackage{setspace}
\usepackage{tabularx}
\usepackage{color}
\usepackage[ colorlinks = true,
             linkcolor = blue,
             urlcolor  = blue,
             citecolor = red,
             anchorcolor = green,
]{hyperref}

\allowdisplaybreaks[2]

\newtheorem{Definition}{Definition}
\newtheorem{Lemma}{Lemma}
\newtheorem{Corollary}[Lemma]{Corollary}
\newtheorem{Proposition}[Lemma]{Proposition}
\newtheorem{Theorem}{Theorem}

\def\Pr{{\rm \bold {Pr}}}

\allowdisplaybreaks[1]

\begin{document}
\flushbottom

%
\title{Strong Converse Theorems for Classes of Multimessage Multicast Networks: A R\'{e}nyi Divergence Approach}
%
%
%


\author{Silas~L.~Fong  and Vincent~Y.~F.~Tan, {\em Senior Member, IEEE} 
\thanks{Silas~L.~Fong and Vincent~Y.~F.~Tan are with the Department of Electrical and Computer Engineering, National University of Singapore (NUS), Singapore (e-mails: \texttt{\{silas\_fong,vtan\}@nus.edu.sg}). Vincent Y.~F.\ Tan is also with the Department of Mathematics, NUS.} \thanks{The authors are supported by NUS grant R-263-000-A98-750/133.}
\thanks{This paper was presented in part at 2015 Information Theory and Applications (ITA) workshop, San Diego, CA, and 2015 IEEE International Symposium on Information Theory (ISIT), Hong Kong. }
}

\maketitle

\begin{abstract}
This paper establishes that the strong converse holds for some classes of discrete memoryless multimessage multicast networks (DM-MMNs) whose corresponding cut-set bounds are tight, i.e., coincide with the set of achievable rate tuples. The strong converse for these classes of DM-MMNs implies that all sequences of codes with rate tuples belonging to the exterior of the cut-set bound have average error probabilities that necessarily tend to one (and are not simply bounded away from zero). Examples in the classes of DM-MMNs
include wireless erasure networks, DM-MMNs consisting of independent discrete memoryless channels (DMCs) as well as single-destination DM-MMNs consisting of independent DMCs with destination feedback. Our elementary proof technique leverages properties of the R\'enyi divergence.
\end{abstract}


\IEEEpeerreviewmaketitle

\section{Introduction}\label{introduction}
\IEEEPARstart{T}{his} paper considers  multimessage multicast networks (MMNs) \cite[Chapter 18]{elgamal} in which the destination nodes want to decode the same set of messages transmitted by the source nodes. A well-known outer bound on the capacity region of the discrete memoryless MMN (DM-MMN) is the {\em cut-set bound}, developed by El Gamal in 1981~\cite{elgamal_81}. This bound states that for any cut $T$ of the network with nodes indexed by $\mathcal{I}$, the sum of the achievable rates of messages on one side of the cut is upper bounded by the conditional mutual information of the input variables in $T$ and the output variables in $T^c \triangleq \mathcal{I}\setminus T$ given the input variables in $T^c$. The DM-MMN is a generalization of the well-studied discrete memoryless relay channel (DM-RC) \cite{CEG}. It is known that the cut-set bound is not tight in general \cite{Aleksic}, but it is tight for several classes of DM-MMNs, including the physically degraded DM-RC~\cite{CEG}, the semi-deterministic DM-RC~\cite{EG82}, the deterministic relay network with no interference \cite{multicastCapacityRelayNetworks}, the finite-field linear deterministic network \cite{AvestimehrDeterministic,linearFiniteField09} and the wireless erasure network \cite{dana06}.

One potential drawback of the cut-set bound is the fact if it is tight, i.e., there exists a matching achievable  inner bound, this only implies a  {\em weak converse} for the problem. In other words, it only guarantees that for all rate tuples \linebreak not belonging to the region prescribed by the cut-set bound, the average error probability in decoding the transmitted  messages is bounded away from zero as the block length of any code tends to infinity. In information theory, it is also important to establish {\em strong converses} as such definitive statements indicate that there is a sharp phase transition between rate tuples that are achievable and those that are not. A strong converse implies that for all codes with rate tuples that are in the exterior of the region prescribed by the fundamental limit, the error probability must necessarily tend to one. The contrapositive of this statement can be stated as follows: All codes whose error probabilities are no larger than $\epsilon\in [0,1)$ as the block length grows, i.e., $\epsilon$-reliable codes, must have rate tuples belonging to the region prescribed by the fundamental limit (in our case, a looser version of the cut-set bound that is tight for some DM-MMNs). This is clearly a stronger statement than the weak converse which considers codes with vanishing error probabilities.

\subsection{Main Contribution}
The main contribution of this work is a self-contained proof of the strong converse for some classes of DM-MMNs in which the cut-set bound is tight. These classes of DM-MMNs include    deterministic relay networks with no interference \cite{multicastCapacityRelayNetworks},   finite-field linear deterministic networks  \cite{AvestimehrDeterministic,linearFiniteField09} and   wireless erasure networks \cite{dana06}.  So for example, for   wireless erasure networks studied by Dana, Gowaiker, Palanki, Hassibi and Effros  \cite{dana06}, all sequences of codes with rates above the capacity   have average error probabilities that necessarily tend to one as the block length grows. The authors of~\cite{dana06} proved using Fano's inequality~\cite[Section~2.10]{CoverBook} that all codes with rates above  capacity have average error probabilities  that are bounded away from zero. Thus,  a consequence of our main result is an important   strengthening of the converse in \cite[Theorem 2]{dana06}.  In addition, we show, using our main theorem, that  the strong converse holds for   \textit{DM-MMNs consisting of independent discrete memoryless channels (DMCs)}  and   \emph{single-destination DM-MMNs consisting of independent DMCs with destination feedback}.  Our main result implies that for the aforementioned DM-MMNs, rate tuples of $\epsilon$-reliable codes where $\epsilon \in [0,1)$ must belong to the region prescribed by the cut-set bound~\cite{elgamal_81}.

The technique that we employ is based on properties of the R\'enyi divergence \cite{yuryRenyiDivergence, renyiDivergencePositivity, cutoff}. This is a powerful technique for establishing strong converses in information theory. It has been employed previously to establish strong converses for point-to-point memoryless DMCs with    output feedback~\cite{arimoto, yuryRenyiDivergence}, classical-quantum channels~\cite{ogawa99} and most recently, entanglement-breaking quantum channels~\cite{wilde14}. We were inspired to use the R\'enyi divergence technique for our strong converse proof because of the similarities of DM-MMNs to channels with full output feedback as shown in the context of sphere-packing bounds on the reliability function for the DM-RC in~\cite{tan14_relay}.
\subsection{Related Work}
The papers that are most closely related to the present work are the ones by Behboodi and Piantanida who   conjectured that the strong converse  holds for DM-RCs~\cite{Beh11} and  general DM multicast  networks~\cite{Beh12}. Also see Appendix~C in the thesis by Behboodi~\cite{Beh_thesis}. It appears to the present authors, however,  that some steps in the justifications, which are based on the information spectrum method \cite{Han10},  are incomplete. Therefore, we are motivated to provide a strong
converse for some (albeit somewhat restrictive) classes of DM-MMNs  using a completely different and elementary method---namely, the R\'enyi divergence approach \cite{yuryRenyiDivergence, renyiDivergencePositivity, cutoff}. As mentioned by Polyanskiy and Verd\'u~\cite{yuryRenyiDivergence}, this approach is arguably the simplest method for proving that memoryless channels with feedback satisfy the strong converse  and thus, we are inspired to leverage it to prove the strong converse   for some classes of DM-MMNs.

\subsection{Paper Outline}
This paper is organized as follows. Section~\ref{notation} presents the notation used in this paper. Section~\ref{sectionDefinition} provides the problem formulation of the DM-MMNs and presents our main theorem. Section~\ref{sectionRenyi} introduces the R\'enyi divergence and discusses its important properties. Section~\ref{sectionSimulatingDistribution} contains an important lemma concerning {\em simulating distributions} which is used in the proof of our main theorem. Section~\ref{sectionMainResult} presents the proof of our main theorem. We also compare and contrast our proof with the proof of the usual cut-set bound which only implies a weak converse. In Section~\ref{sectionMMNwithTightCutSet}, we discuss the above-mentioned classes of DM-MMNs whose cut-set bounds are tight, and we use our main theorem to prove the strong converse  for them. We conclude our discussion and suggest avenues for future research in Section~\ref{sec:conclu}. Proofs of the more technical auxiliary results are relegated to the appendices.

\section{Notation}\label{notation}
We use $\Pr\{\mathcal{E}\}$ to represent the probability of an
event~$\mathcal{E}$, and we let $\boldsymbol{1}(\mathcal{E})$ be the characteristic function of $\mathcal{E}$. We use a capital letter~$X$ to denote a random variable with alphabet $\mathcal{X}$, and use the small letter $x$ to denote a realization of~$X$.
We use $X^n$ to denote a random vector $[X_1\  X_2\  \ldots\  X_n]$, where the components $X_k$ have the same alphabet~$\mathcal{X}$. We let $p_X$ and $p_{Y|X}$ denote the probability mass distribution of $X$ and the conditional probability mass distribution of $Y$ given $X$ respectively for any discrete random variables~$X$ and~$Y$. For any mapping $g$ whose domain includes $\mathcal{X}$, we let $p_{g(X)}$ denote the probability mass distribution of $g(X)$ when $X$ is distributed according to $p_X$. We let $p_X(x)\triangleq\Pr\{X=x\}$ and $p_{Y|X}(y|x)\triangleq\Pr\{Y=y|X=x\}$ be the evaluations of $p_X$ and $p_{Y|X}$ respectively at $X=x$ and $Y=y$. We let $p_Xp_{Y|X}$ denote the joint distribution of $(X,Y)$, i.e., $p_Xp_{Y|X}(x,y)=p_X(x)p_{Y|X}(y|x)$ for all $x$ and $y$. If $X$ and $Y$ are independent, their joint distribution is simply $p_X p_Y$.
 For simplicity, we drop the subscript of a notation if there is no ambiguity. We will take all logarithms to base 2, and we will use the convention that $0\log 0=0$ and $0\log \frac{0}{0}=0$ throughout this paper. For any discrete random variable $(X,Y,Z)$ distributed according to $p_{X,Y,Z}$, we let $H_{p_{X,Z}}(X|Z)$ and $I_{p_{X,Y,Z}}(X;Y|Z)$ be the entropy of $X$ given $Z$ and mutual information between $X$ and $Y$ given $Z$  respectively.
 The $\mathcal{L}_1$-distance between
two distributions $p_X$ and $q_X$ on the same discrete alphabet $\mathcal{X}$, denoted by $\|p_X-q_X\|_{\mathcal{L}_1}$, is defined as
$\|p_X-q_X\|_{\mathcal{L}_1}\triangleq\sum_{x\in\mathcal{X}}|p_X(x)-q_X(x)|$.
  If $X$, $Y$ and $Z$ are distributed according to $p_{X,Y,Z}$ and they form a Markov chain, we write
$(X\rightarrow Y\rightarrow Z)_{p_{X,Y,Z}}$ or more simply, $(X\rightarrow Y\rightarrow Z)_p$.

\section{Problem Formulation and Main Result} \label{sectionDefinition}
We consider a DM-MMN that consists of $N$ nodes. Let
\[
\mathcal{I}\triangleq \{1, 2, \ldots, N\}
\]
be the index set of the nodes, and let $\mathcal{S}\subseteq \mathcal{I}$ and $\mathcal{D}\subseteq \mathcal{I}$ be the sets of sources and destinations respectively. We call $(\mathcal{S}, \mathcal{D})$ the \textit{multicast demand} on the network. The sources in $\mathcal{S}$ transmit information to the destinations in $\mathcal{D}$ in $n$ time slots (channel uses) as follows.
Node~$i$ transmits message
\[
W_{i}\in \{1, 2, \ldots, \lceil 2^{n R_{i}}\rceil\}
 \]
for each $i\in \mathcal{S}$ and node $j$, for each $j \in\mathcal{D}$, wants to decode $\{W_{i}: i\in \mathcal{S}\}$, where $R_{i}$ denotes the rate of message $W_{i}$. We assume that each message $W_{i}$ is uniformly distributed over $\{1, 2, \ldots, \lceil 2^{n R_{i}}\rceil\}$ and all the messages are independent. For each time slot $k\in \{1, 2, \ldots, n\}$ and each $i\in \mathcal{I}$, node~$i$ transmits $X_{i,k} \in \mathcal{X}_i$, a function of $(W_i, Y_i^{k-1})$, and receives, from the output of a channel, $Y_{i,k} \in \mathcal{Y}_i$ where $\mathcal{X}_i$ and $\mathcal{Y}_i$ are some alphabets that  possibly depend on~$i$.
After~$n$ time slots, node~$j$ declares~$\hat W_{i,j}$ to be the
transmitted~$W_{i}$ based on $(W_{j},Y_j^n)$ for each $(i, j)\in \mathcal{S}\times \mathcal{D}$.

To simplify notation, we use the following conventions for each $T\subseteq \mathcal{I}$: For any random tuple \[(X_{1}, X_{2}, \ldots, X_{N}) \in \mathcal{X}_1\times \mathcal{X}_2 \times \ldots \times \mathcal{X}_N,\] we let
 \[
 X_T\triangleq (X_{i}: i\in T)\] be a subtuple of $(X_{1}, X_{2}, \ldots, X_{N})$.
Similarly, for any $k\in \{1, 2, \ldots, n\}$ and any random tuple \[(X_{1,k}, X_{2,k}, \ldots, X_{N, k}) \in \mathcal{X}_1\times \mathcal{X}_2 \times \ldots \times \mathcal{X}_N,\] we let
\[
X_{T,k}\triangleq(X_{i,k}: i\in T)
 \]
 be a subtuple of $(X_{1,k}, X_{2,k}, \ldots, X_{N, k})$. For any $N^2$-dimensional random tuple $(\hat W_{1,1}, \hat W_{1,2}, \ldots, \hat W_{N,N})$, we let
\[
\hat W_{T\times T^c}\triangleq(\hat W_{i,j}: (i,j)\in T\times T^c)
\]
 be a subtuple of $(\hat W_{1,1}, \hat W_{1,2}, \ldots, \hat W_{N,N})$.

The following six definitions formally define a DM-MMN and its capacity region.

\begin{Definition} \label{discreteMemoryless}
A discrete network consists of $N$ finite input sets
$\mathcal{X}_1, \mathcal{X}_2, \ldots, \mathcal{X}_N$, $N$ finite output sets $\mathcal{Y}_1, \mathcal{Y}_2, \ldots, \mathcal{Y}_N$ and a conditional distribution $q_{Y_{\mathcal{I}}|X_{\mathcal{I}}}$. The discrete network is denoted by $(\mathcal{X}_\mathcal{I}, \mathcal{Y}_\mathcal{I}, q_{Y_{\mathcal{I}}|X_{\mathcal{I}}})$.
\end{Definition}


\begin{Definition} \label{defCode}
Let $(\mathcal{X}_\mathcal{I}, \mathcal{Y}_\mathcal{I}, q_{Y_{\mathcal{I}}|X_{\mathcal{I}}})$ be a discrete network, and let $(\mathcal{S}, \mathcal{D})$ be the multicast demand on the network. An $(n, R_\mathcal{I})$-code, where $R_\mathcal{I}$ denotes the tuple of code rates $(R_1, R_2, \ldots, R_N)$, for $n$ uses of the network consists of the following:
\begin{enumerate}
\item A message set
\[
\mathcal{W}_{i}=\{1, 2, \ldots, \lceil 2^{n R_{i}}\rceil\}
\]
 at node~$i$ for each $i\in \mathcal{I}$, where $R_i=0$ for each $i\in \mathcal{S}^c$. Message $W_i$ is uniform on $\mathcal{W}_i$.

\item An encoding function
\[
\phi_{i,k}: \mathcal{W}_{i} \times \mathcal{Y}_i^{k-1} \rightarrow \mathcal{X}_i
 \]
 for each $i\in \mathcal{I}$ and each $k\in\{1, 2, \ldots, n\}$, where $\phi_{i,k}$ is the encoding function at node~$i$ in the
$k^{\text{th}}$ time slot such that
\[
X_{i,k}=\phi_{i,k} (W_{i},
Y_i^{k-1}).
\]
\item A decoding function
\[
\psi_{i,j}: \mathcal{W}_{j} \times
\mathcal{Y}_j^{n} \rightarrow \mathcal{W}_{i}
 \]
 for each $(i, j) \in \mathcal{S}\times \mathcal{D}$, where $\psi_{i,j}$ is the decoding function for message $W_{i}$ at node~$j$ such that
 \[
 \hat W_{i,j} = \psi_{i,j}(W_{j}, Y_j^{n}).
 \]
\end{enumerate}
Since the encoder $\phi_{i,k}$ can depend on the ``feedback signal'' $Y_{i}^{k-1}$, we are allowing full output feedback for each of the transmitting nodes; cf.~Section \ref{sectionWithFeedback}. In addition, the definition of~$\phi_{i,k}$ allows every node to process information in a causal way with a delay of one unit.
\end{Definition}

\begin{Definition}\label{memoryless}
A discrete network $(\mathcal{X}_\mathcal{I}, \mathcal{Y}_\mathcal{I}, q_{Y_{\mathcal{I}}|X_{\mathcal{I}}})$ with multicast demand $(\mathcal{S}, \mathcal{D})$, when used multiple times, is called a \textit{discrete memoryless multimessage multicast network (DM-MMN)} if the following holds for any $(n, R_{\mathcal{I}})$-code:

For all $T\subseteq \mathcal{I}$, we define $q_{Y_{T^c}|X_{\mathcal{I}}}(y_{T^c}| x_{\mathcal{I}})$, the marginal distribution of channel $q_{Y_{\mathcal{I}}|X_{\mathcal{I}}}(y_{\mathcal{I}}| x_{\mathcal{I}})$, as follows:
\begin{equation*}
q_{Y_{T^c}|X_{\mathcal{I}}}(y_{T^c}| x_{\mathcal{I}}) \triangleq  \sum_{y_{T}\in \mathcal{Y}_T}q_{Y_{\mathcal{I}}|X_{\mathcal{I}}}(y_{\mathcal{I}}| x_{\mathcal{I}}) 
\end{equation*}
for all $x_\mathcal{I}\in \mathcal{X}_\mathcal{I}$ and $y_{T^c}\in \mathcal{Y}_{T^c}$. Let $U^{k-1}\triangleq (W_{\mathcal{I}}, X_{\mathcal{I}}^{k-1}, Y_{\mathcal{I}}^{k-1})$ be the collection of random variables that are generated before the $k^{\text{th}}$ time slot. Then, for each $k\in\{1, 2, \ldots, n\}$ and each $T\subseteq \mathcal{I}$,
\begin{align}
 \Pr\{U^{k-1} = u^{k-1}, X_{\mathcal{I},k} =x_{\mathcal{I},k}, Y_{T^c,k}=y_{T^c,k} \}
 = \Pr\{U^{k-1} = u^{k-1}, X_{\mathcal{I},k} =x_{\mathcal{I},k} \} q_{Y_{T^c}|X_{\mathcal{I}}}(y_{T^c,k}| x_{\mathcal{I},k}) \label{memorylessStatement}
\end{align}
for all $u^{k-1}\in \mathcal{U}^{k-1}$, $x_{\mathcal{I},k}\in \mathcal{X}_{\mathcal{I}}$ and $y_{T^c,k}\in \mathcal{Y}_{T^c}$.
\end{Definition}

\begin{Definition} \label{cutseterrorProbability}
For an $(n, R_{\mathcal{I}})$-code defined on the DM-MMN with multicast demand $(\mathcal{S}, \mathcal{D})$,  the {\em average probability of decoding error} is  defined as
\[
\Pr\left\{\bigcup_{j\in \mathcal{D}} \bigcup_{i\in\mathcal{S}} \big\{\hat{W}_{i,j} \ne W_{i } \big\}\right\}.
\]
We call an $(n, R_{\mathcal{I}})$-code with average probability of decoding error not exceeding $\epsilon_n$ an {\em $(n, R_{\mathcal{I}}, \epsilon_n)$-code}.
\end{Definition}

\begin{Definition} \label{cutsetachievable rate}
A rate tuple $R_{\mathcal{I}}$ is \textit{$\epsilon$-achievable} for the DM-MMN with multicast demand $(\mathcal{S}, \mathcal{D})$ if there exists a sequence of $(n, R_{\mathcal{I}}, \epsilon_n)$-codes for the DM-MMN such that
$$ \limsup\limits_{n\rightarrow \infty}\epsilon_n \le \epsilon.$$
\end{Definition}

%
\begin{Definition}\label{cutsetcapacity region}
The \textit{$\epsilon$-capacity region} (for $\epsilon \in [0,1)$) of the DM-MMN with multicast demand $(\mathcal{S}, \mathcal{D})$, denoted by $\mathcal{C}_\epsilon$,  is the set consisting of all  $\epsilon$-achievable rate tuples $R_{\mathcal{I}}$ with $R_{i}=0$ for all $i\in \mathcal{S}^c$. The \textit{capacity region} is defined to be the 0-capacity region $\mathcal{C}_0$.
\end{Definition}

The following theorem is the main result in this paper.

\begin{Theorem} \label{thmMainResult}
Let $(\mathcal{X}_\mathcal{I}, \mathcal{Y}_\mathcal{I}, q_{Y_{\mathcal{I}}|X_{\mathcal{I}}})$ be a DM-MMN with multicast demand $(\mathcal{S}, \mathcal{D})$. Define
\begin{equation}
\mathcal{R}_{\text{out}} \triangleq  \bigcap_{T\subseteq \mathcal{I}: T^c \cap \mathcal{D} \ne \emptyset } \bigcup_{ p_{X_{\mathcal{I}}}} \left\{ R_\mathcal{I}\left| \: \parbox[c]{2.4in}{$ \sum_{ i\in T} R_{i}
 \le  I_{p_{X_{\mathcal{I}}}q_{Y_{T^c}|X_\mathcal{I}}}(X_T; Y_{T^c}|X_{T^c}),\\
 R_i=0 \text{ for all }i\in\mathcal{S}^c$} \right.\right\}. \label{Rout}
\end{equation}
 Then for each $\epsilon \in [0,1)$,
\begin{equation}
\mathcal{C}_\epsilon \subseteq \mathcal{R}_{\text{out}}. \label{strongConverseBound}
\end{equation}
\end{Theorem}
We now make a couple of remarks concerning Theorem~\ref{thmMainResult}.

First, define the usual cut-set bound \cite[Theorem 18.1]{elgamal}
\begin{equation}
\mathcal{R}_{\text{cut-set}}\triangleq  \bigcup_{ p_{X_{\mathcal{I}}}}\bigcap_{T\subseteq \mathcal{I}: T^c \cap \mathcal{D} \ne \emptyset } \left\{ R_\mathcal{I}\left| \: \parbox[c]{2.4in}{$ \sum_{ i\in T} R_{i}
 \le  I_{p_{X_{\mathcal{I}}}q_{Y_{T^c}|X_\mathcal{I}}}(X_T; Y_{T^c}|X_{T^c}),\\
 R_i=0 \text{ for all }i\in\mathcal{S}^c$} \right.\right\}. \label{Rcutset}
\end{equation}
It is well known that $\mathcal{R}_{\text{cut-set}}$ is an outer bound on the capacity region, i.e., that
\begin{equation}
\mathcal{C}_0 \subseteq \mathcal{R}_{\text{cut-set}}. \label{cutSetBound}
\end{equation}
Note that $\mathcal{R}_{\text{out}}$ is similar to $\mathcal{R}_{\text{cut-set}}$ except that the union and the intersection operations are interchanged. Consequently, $\mathcal{R}_{\text{out}}$ is potentially looser (larger) than the $\mathcal{R}_{\text{cut-set}}$. This discrepancy is briefly explained as follows: The proof of Theorem~\ref{thmMainResult}  (i.e., the bound in \eqref{strongConverseBound}) leverages the properties of the R\'{e}nyi divergence, while the proof of the cut-set bound (i.e.,  the bound in \eqref{cutSetBound}) is based on Fano's inequality \cite[Theorem 18.1]{elgamal}. For both proofs, the first step is to fix an achievable rate tuple $R_\mathcal{I}$ and a sequence of $(n, R_\mathcal{I})$-codes. Next a cut $T\subseteq \mathcal{I}$ that satisfies $T^c\cap \mathcal{D}\ne \emptyset$ is also fixed. In both proofs, we eventually arrive at the bound
\begin{equation*}
\sum_{i\in T} R_i \le I_{p_{X_\mathcal{I}}^{(T)}}(X_T; Y_{T^c}|X_{T^c}) 
\end{equation*}
for some $p_{X_\mathcal{I}}^{(T)}$, which implies that
 \begin{equation}
\mathcal{C}_\epsilon \subseteq  \bigcap_{T\subseteq \mathcal{I}: T^c \cap \mathcal{D} \ne \emptyset } \bigcup_{ p_{X_{\mathcal{I}}}^{(T)}} \left\{ R_\mathcal{I}\left| \: \parbox[c]{2.35 in}{$ \sum_{ i\in T} R_{i}
 \le  I_{p_{X_{\mathcal{I}}}^{(T)}q_{Y_{T^c}|X_\mathcal{I}}}(X_T; Y_{T^c}|X_{T^c}),\\
 R_i=0 \text{ for all }i\in\mathcal{S}^c$} \right.\right\}. \label{derivationCutSet*}
\end{equation}
 However, the proofs of bounds \eqref{strongConverseBound} and \eqref{cutSetBound} yield \eqref{derivationCutSet*} under different assumptions on the asymptotic behavior of the average error probability $\epsilon$. For the proof of the cut-set bound \eqref{cutSetBound}, it is assumed that $\epsilon=0$ and hence using Fano's inequality combined with properties of the relative entropy and the conditional mutual information such as the chain rule are sufficient for proving \eqref{derivationCutSet*}. Using Fano's inequality, $p_{X_\mathcal{I}}^{(T)}$ can be shown to be the limit of the sequence of empirical input distributions induced by the sequence of codes (if the limit does not exist, we can always consider a convergent subsequence instead and the following arguments go through in a similar way). In other words,
\begin{align*}
p_{X_\mathcal{I}}^{(T)}(x_\mathcal{I}) =\lim_{n\rightarrow \infty}\frac{1}{n}\sum_{k=1}^n p_{X_{\mathcal{I},k}}(x_\mathcal{I}).
\end{align*}
This implies that $p_{X_{\mathcal{I}}}^{(T)}$ does {\em not} depend on~$T$ and hence the union and the intersection operations in \eqref{derivationCutSet*} can be interchanged, resulting in an improved bound \eqref{cutSetBound}. In contrast, for the proof of our bound \eqref{strongConverseBound}, it is assumed that $\epsilon \in [0,1)$ and hence we need to use properties of the R\'{e}nyi divergence to prove \eqref{derivationCutSet*}. Since $p_{X_{\mathcal{I}}}^{(T)}$ {\em does} depend on~$T$ in general for the proof involving the  R\'{e}nyi divergence, the union and the intersection operations in \eqref{derivationCutSet*} cannot be interchanged in general, which prevents us from further strengthening the bound in \eqref{strongConverseBound}. In Section~\ref{sec:fano}, we   further elaborate  on the similarities of and differences between the proofs of our bound \eqref{strongConverseBound} and the cut-set bound \eqref{cutSetBound}.

Second, although $\mathcal{R}_{\text{out}}$ is potentially looser than the cut-set bound, it can be shown that $\mathcal{R}_{\text{out}} \subseteq \mathcal{C}_\epsilon$ for some classes of networks including the deterministic relay networks with no interference \cite{multicastCapacityRelayNetworks}, the finite-field linear deterministic networks \cite{AvestimehrDeterministic,linearFiniteField09} and the wireless erasure networks \cite{dana06} (discussed in Section~\ref{sectionMulticastNetworks}), the class of DM-MMNs consisting of independent DMCs (discussed in Section~\ref{sectionDM-MMNconsistingOfDMCs}) and the class of single-destination DM-MMNs consisting of independent DMCs with destination feedback (discussed in Section \ref{sectionWithFeedback}). Therefore, Theorem~\ref{thmMainResult} implies the strong converses for these networks.

We briefly outline the content in the sections to follow: The proof of Theorem~\ref{thmMainResult} leverages properties of the R\'{e}nyi divergence, which we   discuss in Section~\ref{sectionRenyi}. In Section~\ref{sectionSimulatingDistribution}, we construct so-called {\em simulating distributions}, which form an important part of the proof of Theorem~\ref{thmMainResult}. The details of the proof of Theorem~\ref{thmMainResult} are provided in Section~\ref{sectionMainResult}. Readers who are only interested in the the application of Theorem \ref{thmMainResult} to specific channel models may proceed directly to Section \ref{sectionMMNwithTightCutSet}.
\section{Properties of the R\'{e}nyi Divergence} \label{sectionRenyi}
The following definitions of (conditional) relative entropy and (conditional) R\'{e}nyi divergence are standard~\cite{yuryRenyiDivergence, renyiDivergencePositivity, cutoff}.

\begin{Definition}\label{defRenyiDivergence}
Let $p_{X}$ and $q_{X}$ be two probability distributions on $\mathcal{X}$, and let $r_Z$ be a probability distribution on $\mathcal{Z}$.
Let
\[
D(p_X\|q_X)\triangleq \sum_{x\in\mathcal{X}}p_X(x)\log\frac{p_X(x)}{q_X(x)}
 \]
 be the {\em relative entropy} between $p_X$ and $q_X$, and let
\begin{align*}
D(p_{X|Z}\|q_{X|Z}|r_Z)\triangleq \sum_{z\in\mathcal{Z}}r_Z(z) D(p_{X|Z=z}\|q_{X|Z=z})
 \end{align*}
 be the {\em conditional relative entropy} between $p_{X|Z}$ and $q_{X|Z}$ conditioned on $r_Z$.
Then, the {\em R\'{e}nyi divergence with parameter} $\lambda\in [1, \infty)$ between $p_{X}$ and $q_{X}$, denoted by $D_\lambda(p_X\|q_X)$, is defined as follows:
\begin{align*}
 D_\lambda(p_X\|q_X) \triangleq
 \begin{cases}
 \frac{1}{\lambda-1}\log\sum\limits_{x\in\mathcal{X}}\frac{(p_X(x))^\lambda}{(q_X(x))^{\lambda-1}} &\text{if $\lambda>1$,}\\
 D(p_X\|q_X)  &\text{if $\lambda=1$}.
 \end{cases}
\end{align*}
In addition, the {\em conditional R\'{e}nyi  divergence with parameter} $\lambda\in [1, \infty)$ between $p_{X|Z}$ and $q_{X|Z}$ given $r_Z$, denoted by $D_\lambda(p_{X|Z}\|q_{X|Z}|r_{Z})$, is defined as follows:
\begin{align*}
 D_\lambda(p_{X|Z}\|q_{X|Z}|r_Z) & \triangleq
\begin{cases}
\frac{1}{\lambda-1} \log \sum\limits_{z\in\mathcal{Z}} r_Z(z) \sum\limits_{x\in\mathcal{X}}\frac{(p_{X|Z}(x|z))^\lambda}{(q_{X|Z}(x|z))^{\lambda-1}} &\text{if $\lambda>1$,}
\\
 D(p_{X|Z}\|q_{X|Z}|r_Z) &\text{if $\lambda=1$.}
 \end{cases}
\end{align*}
 Note that for $\lambda>1$, $D_\lambda(p_{X|Z}\|q_{X|Z}|r_Z)$ can be expressed in terms of the unconditional R\'{e}nyi divergence as
\[
D_\lambda(p_{X|Z}\|q_{X|Z}|r_Z) = \frac{1}{\lambda-1}\log\sum\limits_{z\in\mathcal{Z}} r_Z(z) 2^{(\lambda-1)D_\lambda (p_{X|Z=z}\| q_{X|Z=z})}.
\]
\end{Definition}

We summarize two important properties of $D_\lambda(p_{X|Z}\|q_{X|Z}|r_{Z})$ in the following theorem, whose proof can be found in \cite[Theorems 5 and 9]{renyiDivergence}.

\begin{Theorem}\label{thmRenyiDivergence} For any $\lambda\in [1, \infty)$, the following statements hold for any two conditional  probability distributions $p_{X,Y|Z}$, $q_{X,Y|Z}$ and any probability distribution $r_Z$:
\begin{itemize}
\item[1.] (Continuity)  $D_\lambda(p_{X|Z}\|q_{Y|Z}|r_Z)$ is continuous in $\lambda$.
  \item[2.] (Data processing inequality (DPI))
$D_\lambda(p_X\|q_X) \ge D_\lambda(p_{g(X)}\|q_{g(X)})$ for any function $g$  with domain $\mathcal{X}$. In particular, $D_\lambda(p_{X,Y}\|q_{X,Y}) \ge D_\lambda(p_{X}\|q_{X})$.
\end{itemize}
\end{Theorem}

Most converse theorems use Fano's inequality \cite[Section~2.10]{Cov06} to obtain a lower bound on the error probability. However, this can only lead to weak converse results. The following proposition, analogous to Fano's inequality, enables us to prove strong converse results by providing a better lower bound on the error probability. Essentially, we have the freedom to choose any $\lambda \in (1,\infty)$ in the bound  in~\eqref{propositionDlambdaLowerBoundEq1} below.

\begin{Proposition} \label{propositionDlambdaLowerBound}
Let $p_{U,V}$ be a probability distribution defined on $\mathcal{W}\times \mathcal{W}$ for some $\mathcal{W}$, and let $p_U$ be the marginal distribution of $p_{U,V}$. In addition, let $q_{V}$ be a distribution defined on $\mathcal{W}$. Suppose $p_{U}$ is the uniform distribution, and let
\begin{equation}
\alpha = \Pr\{U\ne V\} \label{defEpsilon}
\end{equation}
be a real number in $[0, 1)$. Then for each $\lambda\in (1, \infty)$,
\begin{equation}
D_\lambda(p_{U,V}\|p_Uq_{V}) \ge \log|\mathcal{W}| + \lambda(\lambda-1)^{-1}\log(1-\alpha). \label{propositionDlambdaLowerBoundEq1}
\end{equation}
\end{Proposition}
\begin{IEEEproof}
  Fix a $\lambda\in (1, \infty)$ and let $s_{U,V}\triangleq p_Uq_V$. Consider the following chain of inequalities:
\begin{align*}
 D_\lambda(p_{U,V}\|s_{U,V})
 & \stackrel{\text{(a)}}{\ge} D_\lambda(p_{\boldsymbol{1}(\{U=V\})}\|s_{\boldsymbol{1}(\{U=V\})}) \notag \\
&  \stackrel{\text{(b)}}{=} \frac{1}{\lambda-1}\log\left(|\mathcal{W}|^{\lambda-1}(1-\alpha)^\lambda + \left(\frac{|\mathcal{W}|}{|\mathcal{W}|-1}\right)^{\lambda-1}\alpha^\lambda\right) \notag \\
&\stackrel{\text{(c)}}{\ge} \log|\mathcal{W}| + \lambda(\lambda-1)^{-1}\log(1-\alpha), 
\end{align*}
where
\begin{enumerate}
\item[(a)] follows from the DPI in Theorem~\ref{thmRenyiDivergence};
\item[(b)] follows from Definition~\ref{defRenyiDivergence} and the facts that
\begin{align*}
\sum_{u,v}p_{U,V}(u,v)\boldsymbol{1}(\{u=v\}) &= 1- \alpha \quad \mbox{(cf.\ \eqref{defEpsilon})}   \quad\mbox{and}\\
\sum_{u,v}s_{U,V}(u,v)\boldsymbol{1}(\{u=v\}) &= \frac{1}{|\mathcal{W}|};
\end{align*}
\item[(c)] follows from the fact that $\big(\frac{|\mathcal{W}|}{|\mathcal{W}|-1}\big)^{\lambda-1}\alpha^\lambda\ge 0$.
\end{enumerate}
This completes the proof.
\end{IEEEproof}

The following proposition enables us to approximate the conditional R\'enyi divergence $D_\lambda$ by the conditional relative entropy $D_1=D$. Since the proof for the following proposition is straightforward but involves some tedious algebra, we defer it to Appendix~\ref{appendixA}.
\begin{Proposition} \label{propositionDlambdaToMutualInfo}
Let $\lambda\in [1, 5/4]$ be a real number, and let $p_{X,Y,Z}$ be a probability distribution defined on $\mathcal{X}\times \mathcal{Y}\times \mathcal{Z}$. Then,
\begin{equation}
D_\lambda(p_{X,Y|Z}||p_{X|Z}p_{Y|Z}|p_Z) \le D(p_{X,Y|Z}||p_{X|Z}p_{Y|Z}|p_Z) +8(\lambda-1) (|\mathcal{X}||\mathcal{Y}|)^{5}. \label{propositionDlambdaToMutualInfoStatement}
\end{equation}
\end{Proposition}
We made no attempt to optimize the remainder term  $8 (|\mathcal{X}||\mathcal{Y}|)^{5}$ as the important part of the statement is that this remainder term is uniform in $p_{X,Y,Z}$ on a sufficiently small interval to the right of $\lambda=1$. In fact, it only depends on the product $|\mathcal{X}||\mathcal{Y}|$.

\section{Simulating Distribution} \label{sectionSimulatingDistribution}
Proposition~\ref{propositionDlambdaLowerBound} provides a lower bound for the error probability, and the lower bound holds for all $q_V$. Therefore, we are motivated to choose a \textit{simulating distribution} $q_V$ so that the left hand side of \eqref{propositionDlambdaLowerBoundEq1} can be simplified. Before describing the simulating distribution, we state the following proposition which facilitates to characterize an important property of  Markov chains. 

\begin{Proposition} \label{propositionMCsimplification}
Suppose there exist two probability distributions $r_{X,Y}$ and $q_{Z|Y}$ such that
\begin{equation}
p_{X, Y, Z}(x,y,z) = r_{X,Y}(x,y)q_{Z|Y}(z|y) \label{statement1CorollaryMC}
\end{equation}
for all $x$, $y$ and $z$ whenever $p_Y(y)>0$. Then
\begin{equation}
(X\rightarrow Y\rightarrow Z)_{p_{X,Y,Z}} \label{statement3CorollaryMC}
\end{equation}
forms a Markov chain. In addition,
\begin{equation}
p_{Z|Y}=q_{Z|Y} \label{statement2CorollaryMC}.
\end{equation}
\end{Proposition}
\begin{IEEEproof}The proof of \eqref{statement3CorollaryMC} is contained \cite[Proposition~2.5]{Yeung08Book}. It remains to show \eqref{statement2CorollaryMC}. Summing $x$ and then $z$ on both sides of \eqref{statement1CorollaryMC}, we have $p_{Y,Z}(y,z)=r_Y(y)q_{Z|Y}(z|y)$ and $p_Y(y) = r_Y(y)$ for all $x$, $y$ and $z$ whenever $p_Y(y)>0$, which implies \eqref{statement2CorollaryMC}.
\end{IEEEproof}

The construction of the simulating distribution is contained in the following lemma. Before stating lemma, we make the following definitions: Given an $(n, R_{\mathcal{I} }, \epsilon_n)$-code, we let $p_{W_{\mathcal{I}},X_\mathcal{I}^n, Y_\mathcal{I}^n, \hat W_{\mathcal{I}\times \mathcal{I}}}$ be the probability distribution induced by the code according to Definitions~\ref{defCode} and~\ref{memoryless}. In the following, we drop the subscripts of the probability distributions to simplify notation. For each $T\subseteq\mathcal{I}$ and each $\lambda\in[1, \infty)$, recalling that $q_{Y_{T^c}|X_{\mathcal{I}}}$ denotes the channel of the DM-MMN defined in Definition~\ref{memoryless}, we define $s_{X_{\mathcal{I},1}, Y_{T^c,1} }^{(\lambda, T)}\triangleq  p_{X_{\mathcal{I},1}}q_{Y_{T^c}|X_\mathcal{I}}$. Then, we define $s_{X_{\mathcal{I},k}}^{(\lambda, T)}$ and $s_{X_{\mathcal{I},k}, Y_{T^c,k}}^{(\lambda, T)}$ for $k=2, \ldots, n$ based on $\{s_{X_{\mathcal{I},\ell}, Y_{T^c,\ell}}^{(\lambda, T)}\}_{\ell=1}^{k-1}$ as follows: For all $x_{\mathcal{I},k}\in \mathcal{X}_\mathcal{I}$ and $y_{T^c,k}\in \mathcal{Y}_{T^c}$, let
\begin{align}
 s_{X_{\mathcal{I},k}}^{(\lambda, T)}(x_{\mathcal{I},k}) \triangleq \frac{  \sum_{x_{\mathcal{I}}^{k-1}, y_{T^c}^{k-1}} p(x_{\mathcal{I},k}|  x_{\mathcal{I}}^{k-1},   y_{T^c}^{k-1}) \prod_{\ell=1}^{k-1} \left(p(x_{\mathcal{I},\ell}| x_{\mathcal{I}}^{\ell-1}, y_{T^c}^{\ell-1}) \frac{\left(q(y_{T^c,\ell}| x_{\mathcal{I},\ell})\right)^\lambda}{\left(s^{(\lambda, T)}(y_{T^c,\ell}| x_{T^c,\ell})\right)^{\lambda-1}} \right) }{\sum_{x_{\mathcal{I}}^{k-1}, y_{T^c}^{k-1}}\prod_{\ell=1}^{k-1} \left(p(x_{\mathcal{I},\ell}| x_{\mathcal{I}}^{\ell-1}, y_{T^c}^{\ell-1}) \frac{\left(q(y_{T^c,\ell}| x_{\mathcal{I},\ell})\right)^\lambda}{\left(s^{(\lambda, T)}(y_{T^c,\ell}| x_{T^c,\ell})\right)^{\lambda-1}}\right)} \label{defsRecursive}
\end{align}
and
 \begin{equation}
s_{X_{\mathcal{I},k},Y_{T^c,k}}^{(\lambda, T)}(x_{\mathcal{I},k},y_{T^c,k}) \triangleq
s_{X_{\mathcal{I},k}}^{(\lambda, T)}(x_{\mathcal{I},k})
q_{Y_{T^c}|X_{\mathcal{I}}}(y_{T^c,k}|x_{\mathcal{I},k}).
\label{defsRecursive**}
 \end{equation}
It can be verified by using \eqref{memorylessStatement}, \eqref{defsRecursive} and \eqref{defsRecursive**} that $s_{X_{\mathcal{I},k}, Y_{T^c,k}}^{(1, T)}=p_{X_{\mathcal{I},k}, Y_{T^c,k}}$, and hence $s_{X_{\mathcal{I},k}, Y_{T^c,k}}^{(\lambda, T)}$ can be viewed as a tilted version of $p_{X_{\mathcal{I},k}, Y_{T^c,k}}$. More specifically, we can see from \eqref{defsRecursive} that $s_{X_{\mathcal{I},k}}^{(\lambda, T)}$ can be viewed as a weighted version of $p_{X_{\mathcal{I},k}|  X_{\mathcal{I}}^{k-1}, Y_{T^c}^{k-1}}$ where the weighting distribution is a tilting of $\prod_{\ell=1}^{k-1}(p_{X_{\mathcal{I},\ell}|  X_{\mathcal{I}}^{\ell-1}, Y_{T^c}^{\ell-1}}q_{Y_{T^c}| X_{\mathcal{I}}})$ towards $\prod_{\ell=1}^{k-1}(p_{X_{\mathcal{I},\ell}|  X_{\mathcal{I}}^{\ell-1}, Y_{T^c}^{\ell-1}}s_{Y_{T^c,\ell}| X_{T^c,\ell}}^{(\lambda, T)})$.
\medskip

\begin{Lemma} \label{lemmaSimulatingDistribution}
 Given an $(n, R_{\mathcal{I} }, \epsilon_n)$-code for the DM-MMN, let $p_{W_{\mathcal{I}},X_\mathcal{I}^n, Y_\mathcal{I}^n, \hat W_{\mathcal{I}\times \mathcal{I}}}$ be the probability distribution induced by the code according to Definitions~\ref{defCode} and~\ref{memoryless}. Let $T$ be an arbitrary subset of $\mathcal{I}$ and fix an arbitrary $\lambda\in[1, \infty)$. Then there exists a probability distribution $s_{W_{\mathcal{I}},X_\mathcal{I}^n, Y_\mathcal{I}^n, \hat W_{\mathcal{I}\times \mathcal{I}}}$ that satisfies the following properties:
 \begin{enumerate}
 \item[(i)] $s_{W_{\mathcal{I}}} = p_{W_{\mathcal{I}}}$.
 \item[(ii)] $s_{\hat W_{T \times T^c}|W_{T^c},Y_{T^c}^n} = p_{\hat W_{T \times T^c}|W_{T^c},Y_{T^c}^n}$.
 \item [(iii)] For each $k\in\{1, 2, \ldots, n\}$, $\left((W_{\mathcal{I}},X_{T^c}^{k-1},Y_{T^c}^{k-1})\rightarrow X_{T^c,k} \rightarrow Y_{T^c,k}\right)_{s}$ forms a Markov chain.
     \item[(iv)] For each $k\in\{1, 2, \ldots, n\}$, $s_{Y_{T^c,k}|X_{T^c,k}} = s_{Y_{T^c,k}|X_{T^c,k}}^{(\lambda, T)}$, where $s_{Y_{T^c,k}|X_{T^c,k}}^{(\lambda, T)}$ is induced by the joint distribution in \eqref{defsRecursive**}.
 \item[(v)] For each $k\in\{1, 2, \ldots, n\}$, $p_{X_{T^c,k}|W_{\mathcal{I}}, X_{\mathcal{I}}^{k-1}, Y_{T^c}^{k-1}}=s_{X_{T^c,k}|W_{T^c}, X_{T^c}^{k-1} Y_{T^c}^{k-1}}$.
 \end{enumerate}
 We call $s_{W_{\mathcal{I}},X_\mathcal{I}^n, Y_\mathcal{I}^n, \hat W_{\mathcal{I}\times \mathcal{I}}}$ a \textit{$\lambda$-simulating distribution of $p_{W_{\mathcal{I}},X_\mathcal{I}^n, Y_\mathcal{I}^n, \hat W_{\mathcal{I}\times \mathcal{I}}}$ neglecting $T$} because $s_{X_{\mathcal{I},k}, Y_{T^c,k}}$ represents a ``$\lambda$-tilting" of $p_{X_{\mathcal{I},k}, Y_{T^c,k}}$ through Property~(iv) and captures all the important properties of $(X_{T^c}^n, Y_{T^c}^n)$ when $(X_{T^c}^n, Y_{T^c}^n)$ is generated according to the given code distribution  $p_{W_{\mathcal{I}},X_\mathcal{I}^n, Y_\mathcal{I}^n, \hat W_{\mathcal{I}\times \mathcal{I}}}$.
\end{Lemma}
\begin{IEEEproof}
We prove the lemma by first constructing a distribution of $(W_{\mathcal{I}},X_{T^c}^n, Y_{T^c}^n)$ denoted by $r$. Subsequently, we use $r$ as a building block to construct a distribution of $(W_{\mathcal{I}},X_{\mathcal{I}}^n, Y_{\mathcal{I}}^n, \hat W_{\mathcal{I}\times \mathcal{I}})$.
Define
\begin{equation}
r_{W_{\mathcal{I}},X_{T^c}^1, Y_{T^c}^1} \triangleq p_{W_{\mathcal{I}}} p_{X_{T^c,1}|W_{T^c}}s_{Y_{T^c,1}|X_{T^c,1}}^{(\lambda, T)}. \label{rConstructionFirstStep}
\end{equation}
Recursively construct
\begin{equation}
r_{W_{\mathcal{I}},X_{T^c}^k, Y_{T^c}^k}  \triangleq
r_{W_{\mathcal{I}},X_{T^c}^{k-1}, Y_{T^c}^{k-1}}  p_{X_{T^c,k}|W_{T^c}, X_{T^c}^{k-1}, Y_{T^c}^{k-1}}  s_{Y_{T^c,k}|X_{T^c,k}}^{(\lambda, T)} \label{qSimulatingConstructionRecursively}
 \end{equation}
for each $k=2,3,\ldots, n$, where $s_{Y_{T^c,k}|X_{T^c,k}}^{(\lambda, T)} $ is as defined in \eqref{defsRecursive**}.
Applying \eqref{qSimulatingConstructionRecursively} recursively from $k=2$ to $k=n$ and using \eqref{rConstructionFirstStep}, we have
\begin{equation}
r_{W_{\mathcal{I}},X_{T^c}^n, Y_{T^c}^n}  = p_{W_{\mathcal{I}}} \prod_{k=1}^n \left( p_{X_{T^c,k}|W_{T^c}, X_{T^c}^{k-1}, Y_{T^c}^{k-1}} s_{Y_{T^c,k}|X_{T^c,k}}^{(\lambda, T)}\right).\label{rConstructionOverall}
\end{equation}
After defining $r$ through \eqref{rConstructionFirstStep}, \eqref{qSimulatingConstructionRecursively} and \eqref{rConstructionOverall}, we are now ready to define $s$ as follows:
\begin{equation}
s_{W_{\mathcal{I}},X_\mathcal{I}^n, Y_\mathcal{I}^n, \hat W_{\mathcal{I}\times \mathcal{I}}} \triangleq
p_{X_{T}^n,Y_{T}^n,\hat W_{(T\times T^c)^c}}r_{W_{\mathcal{I}},X_{T^c}^n, Y_{T^c}^n} p_{\hat W_{T\times T^c}|W_{T^c},Y_{T^c}^n}. \label{qSimulatingConstruction}
 \end{equation}
In the rest of the proof, we want to show that $s_{W_{\mathcal{I}},X_\mathcal{I}^n, Y_\mathcal{I}^n, \hat W_{\mathcal{I}\times \mathcal{I}}}$ satisfies Properties (i), (ii), (iii), (iv) and (v).

Since
\begin{align*}
&\sum_{x_\mathcal{I}^n, y_\mathcal{I}^n, {\hat w}_{\mathcal{I}\times \mathcal{I}}} s_{W_{\mathcal{I}},X_\mathcal{I}^n, Y_\mathcal{I}^n, \hat W_{\mathcal{I}\times \mathcal{I}}}(w_{\mathcal{I}}, x_\mathcal{I}^n, y_\mathcal{I}^n, {\hat w}_{\mathcal{I}\times \mathcal{I}})\notag\\
& \stackrel{\eqref{qSimulatingConstruction}}{=} \sum_{x_{T^c}^n,y_{T^c}^n} r_{W_{\mathcal{I}},X_{T^c}^n, Y_{T^c}^n}(w_{\mathcal{I}},x_{T^c}^n, y_{T^c}^n)
\notag \\ &\stackrel{\eqref{rConstructionOverall}}{=}p_{W_{\mathcal{I}}}(w_{\mathcal{I}})
\end{align*}
for all $w_{\mathcal{I}}$, it follows that Property (i) holds.

In order to prove Property (ii), we write
\begin{align*}
s_{W_{T^c},Y_{T^c}^n,\hat W_{T\times T^c}}\stackrel{\eqref{qSimulatingConstruction}}{=} r_{W_{T^c}, Y_{T^c}^n} p_{\hat W_{T\times T^c}|W_{T^c},Y_{T^c}^n}
\end{align*}
which implies from Proposition~\ref{propositionMCsimplification} that $s_{\hat W_{T\times T^c}|W_{T^c},Y_{T^c}^n}=p_{\hat W_{T\times T^c}|W_{T^c},Y_{T^c}^n}$.

In order to prove Properties (iii), (iv) and (v), we write for each $k\in\{1,2,\ldots, n\}$
\begin{align}
s_{W_{\mathcal{I}}, X_{T^c}^k, Y_{T^c}^k}
 &\stackrel{\eqref{qSimulatingConstruction}}{=} r_{W_{\mathcal{I}},X_{T^c}^k, Y_{T^c}^k} \notag\\*
 & \stackrel{\text{(a)}}{=} p_{W_{\mathcal{I}}} \prod_{m=1}^k \left(p_{X_{T^c,m}|W_{T^c}, X_{T^c}^{m-1}, Y_{T^c}^{m-1}} s_{Y_{T^c,m}|X_{T^c,m}}^{(\lambda, T)}\right), \label{eqnProperty3}
\end{align}
where (a) follows from marginalizing \eqref{rConstructionOverall}. It then follows from \eqref{eqnProperty3} and Proposition~\ref{propositionMCsimplification} that for each $k\in\{1, 2, \ldots, n\}$,
\begin{equation}
\left((W_{\mathcal{I}}, X_{T^c}^{k-1}, Y_{T^c}^{k-1})\rightarrow X_{T^c,k} \rightarrow Y_{T^c,k} \right)_{s} \label{rDistributionTemp2}
\end{equation}
forms a Markov chain and
\begin{equation}
s_{ Y_{T^c,k}|X_{T^c,k}}= s_{Y_{T^c,k}|X_{T^c,k}}^{(\lambda, T)}. \label{rDistributionTemp3}
\end{equation}
Properties~(iii) and~(iv) follow from \eqref{rDistributionTemp2} and \eqref{rDistributionTemp3} respectively.
In addition, for each $k\in\{1, 2, \ldots, n\}$,
   \begin{align}
s_{W_{\mathcal{I}},X_{T^c}^k, Y_{T^c}^{k-1}} &  \stackrel{\eqref{eqnProperty3}}{=}   p_{W_{\mathcal{I}}} p_{X_{T^c,k}|W_{T^c}, X_{T^c}^{k-1}, Y_{T^c}^{k-1}}  \prod_{m=1}^{k-1} \left(p_{X_{T^c,m}|W_{T^c}, X_{T^c}^{m-1}, Y_{T^c}^{m-1}} s_{Y_{T^c,m}|X_{T^c,m}}^{(\lambda, T)}\right).
   \label{rDistributionTemp4}
 \end{align}
 Then, for each $k\in\{1, 2, \ldots, n\}$,
\begin{align}
 p_{X_{T^c,k}|W_{\mathcal{I}}, X_{\mathcal{I}}^{k-1}, Y_{T^c}^{k-1}}  &\stackrel{\text{(a)}}{=} p_{X_{T^c,k}|W_{T^c}, X_{T^c}^{k-1}, Y_{T^c}^{k-1}} \notag\\
 &\stackrel{\text{(b)}}{=}  s_{X_{T^c,k}|W_{T^c},X_{T^c}^{k-1},Y_{T^c}^{k-1}} \label{rDistributionTemp1}
\end{align}
 where
 \begin{enumerate}
 \item[(a)] follows from the fact that $( (W_{T}, X_\mathcal{I}^{k-1}) \rightarrow (W_{T^c},Y_{T^c}^{k-1}) \rightarrow X_{T^c,k})_p$
 forms a Markov chain (cf.\ Definition~\ref{defCode}).
 \item[(b)] follows from \eqref{rDistributionTemp4} and Proposition~\ref{propositionMCsimplification}.
 \end{enumerate}
Property~(v) follows from \eqref{rDistributionTemp1}.
\end{IEEEproof}
\section{Proof of Theorem \ref{thmMainResult}} \label{sectionMainResult}
We partition the proof into several subsections for the sake of clarity and readability. In the final subsection (Section~\ref{sec:fano}), we compare and contrast the proof of Theorem \ref{thmMainResult} with the proof of the usual cut-set bound which only implies a weak converse.
\subsection{Lower Bounding the Error Probability in Terms of the R\'{e}nyi Divergence} \label{subsectionLowerBound}
Fix an $\epsilon\in [0,1)$ and let $R_{\mathcal{I}}$ be an $\epsilon$-achievable rate tuple for the DM-MMN. By Definitions~\ref{cutsetachievable rate} and~\ref{cutsetcapacity
region}, there exists a number $\bar \epsilon\in [0,1)$ and a sequence of $(n, R_{\mathcal{I}}, \epsilon_n)$-codes on the DM-MMN such that for all sufficiently large~$n$,
\begin{equation}
 \epsilon_n \le \bar \epsilon. \label{thmTempEq2}
\end{equation}
Fix a sufficiently large~$n$ such that \eqref{thmTempEq2} holds, and let $p_{W_{\mathcal{I}},X_\mathcal{I}^n, Y_\mathcal{I}^n, \hat W_{\mathcal{I}\times \mathcal{I}}}$ be the probability distribution induced by the $(n, R_{\mathcal{I}}, \epsilon_n)$-code on the DM-MMN. Fix an arbitrary $T\subseteq \mathcal{I}$ such that $T^c \cap \mathcal{D} \ne \emptyset$, and choose a node $d\in T^c \cap \mathcal{D}$. Fix an arbitrary $\lambda\in (1, \infty)$.  Let $s_{W_{\mathcal{I}},X_\mathcal{I}^n, Y_\mathcal{I}^n, \hat W_{\mathcal{I}\times \mathcal{I}}}$ be a $\lambda$-simulating distribution of $p_{W_{\mathcal{I}},X_\mathcal{I}^n, Y_\mathcal{I}^n, \hat W_{\mathcal{I}\times \mathcal{I}}}$ neglecting~$T$ such that $s_{W_{\mathcal{I}},X_\mathcal{I}^n, Y_\mathcal{I}^n, \hat W_{\mathcal{I}\times \mathcal{I}}}$ satisfies all the properties in Lemma~\ref{lemmaSimulatingDistribution}. Then, it follows from Proposition~\ref{propositionDlambdaLowerBound} and Definition~\ref{defCode} with the identifications $U\equiv W_{T}$, $V\equiv \hat W_{T\times \{d\}}$, $p_{U,V}\equiv  p_{W_{T},  \hat W_{T\times \{d\}}}$, $q_V\equiv  s_{\hat W_{T\times \{d\}}}$, $|\mathcal{W}|\equiv  2^{\sum_{i\in T}\lceil nR_{i}\rceil}$ and $\alpha\equiv \Pr\{W_{T}\ne  \hat W_{T\times \{d\}}\} \le \bar \epsilon$ that
 \begin{align}
& D_\lambda(p_{W_{T}, \hat W_{T\times \{d\}}}\|p_{W_{T}}s_{\hat W_{T\times \{d\}}}) \nonumber\\
  &\ge \sum_{i\in T}nR_i + \lambda(\lambda-1)^{-1}\log(1-\alpha) \nonumber\\
&  \ge\sum_{i\in T}nR_i + \lambda(\lambda-1)^{-1}\log(1-\bar \epsilon). \label{eqnDLambdaReversechain}
 \end{align}
 \subsection{Using the DPI to Introduce the Channel Input and Output}
Let $U_{T^c}^{k-1}\triangleq (W_{\mathcal{I}}, X_{\mathcal{I}}^{k-1},Y_{T^c}^{k-1})$ and $V_{T^c}^{k-1}\triangleq (W_{T^c},X_{T^c}^{k-1},Y_{T^c}^{k-1})$ be the random variables generated before the $k^{\text{th}}$ time slot, and consider the following chain of inequalities:
 \begin{align}
& D_\lambda(p_{W_{T},\hat W_{T\times \{d\}}}\|p_{W_{T}} s_{\hat W_{T\times \{d\}}}) \notag\\
&\stackrel{\text{(a)}}{\le} D_\lambda(p_{U_{T^c}^{0},\hat W_{T\times \{d\}}}\|p_{W_{T}} s_{V_{T^c}^{0},\hat W_{T\times \{d\}}}) \notag\\
&= D_\lambda(p_{U_{T^c}^{0},\hat W_{T\times \{d\}}}\|p_{W_{T}} s_{V_{T^c}^{0}}s_{\hat W_{T\times \{d\}}|V_{T^c}^{0}}) \notag\\
&\stackrel{\text{(b)}}{=} D_\lambda(p_{U_{T^c}^{0},\hat W_{T\times \{d\}}}\|p_{W_{T}} p_{V_{T^c}^{0}}s_{\hat W_{T\times \{d\}}|V_{T^c}^{0}}) \notag\\
&\stackrel{\text{(c)}}{=}D_\lambda(p_{U_{T^c}^{0}}p_{\hat W_{T\times \{d\}}|U_{T^c}^{0}}\|p_{U_{T^c}^{0}} s_{\hat W_{T\times \{d\}}|V_{T^c}^{0}}) \notag\\
&\stackrel{\text{(d)}}{\le} D_\lambda(p_{U_{T^c}^{0}}p_{\hat W_{T\times \{d\}}, Y_{T^c}^n|U_{T^c}^{0}}\|p_{U_{T^c}^{0}} s_{\hat W_{T\times \{d\}}, Y_{T^c}^n|V_{T^c}^{0}}) \notag\\
& = D_\lambda(p_{U_{T^c}^{0}}p_{Y_{T^c}^n|U_{T^c}^{0}}p_{\hat W_{T\times \{d\}}| U_{T^c}^{0},Y_{T^c}^n}\|p_{U_{T^c}^{0}} s_{Y_{T^c}^n|V_{T^c}^{0}} s_{\hat W_{T\times \{d\}}| V_{T^c}^{0},Y_{T^c}^n}) \notag\\
& \stackrel{\text{(e)}}{=} D_\lambda(p_{U_{T^c}^{0}}p_{\hat W_{T\times \{d\}}| V_{T^c}^{0},Y_{T^c}^n}p_{Y_{T^c}^n|U_{T^c}^{0}}\|p_{U_{T^c}^{0}}  p_{\hat W_{T\times \{d\}}| V_{T^c}^{0},Y_{T^c}^n}s_{Y_{T^c}^n|V_{T^c}^{0}}) \notag\\
&\stackrel{\text{(f)}}{\le} D_\lambda\left(p_{U_{T^c}^{0}}p_{\hat W_{T\times \{d\}}| V_{T^c}^{0},Y_{T^c}^n}p_{X_{\mathcal{I}}^n,Y_{T^c}^n|U_{T^c}^{0}}\left\|p_{U_{T^c}^{0}}p_{\hat W_{T\times \{d\}}| V_{T^c}^{0},Y_{T^c}^n} s_{X_{T^c}^n,Y_{T^c}^n|V_{T^c}^{0}}\prod_{k=1}^n p_{X_{T,k}|U_{T^c}^{k-1}, X_{T^c,k} } \right.\right) \label{eqnDlambdaBeforeFirstChain}
\end{align}
where
 \begin{enumerate}
 \item[(a)] follows from the DPI of $D_\lambda$ by introducing $W_{T^c}$.
 \item[(b)] follows from Property~(i) in Lemma~\ref{lemmaSimulatingDistribution}.
 \item[(c)] follows from the fact that $W_{T}$ and $W_{T^c}$ are independent.
 \item[(d)] follows from the DPI of $D_\lambda$ by introducing the channel output $Y_{T^c}^n$.
 \item[(e)] follows from Property~(ii) in Lemma~\ref{lemmaSimulatingDistribution} and the fact that
 \[
 ( W_{T}  \rightarrow (W_{T^c},Y_{T^c}^n) \rightarrow \hat W_{T\times \{d\}})_p
 \]
 forms a Markov chain.
 \item[(f)] follows from the DPI of $D_\lambda$ by introducing the channel input $X_\mathcal{I}^n$.
 \end{enumerate}
 In order to simplify \eqref{eqnDlambdaBeforeFirstChain}, we consider
\begin{align}
&D_\lambda\left(p_{U_{T^c}^{0}}p_{\hat W_{T\times \{d\}}| V_{T^c}^{0},Y_{T^c}^n}p_{X_{\mathcal{I}}^n,Y_{T^c}^n|U_{T^c}^{0}}\left\|p_{U_{T^c}^{0}}p_{\hat W_{T\times \{d\}}| V_{T^c}^{0},Y_{T^c}^n} s_{X_{T^c}^n,Y_{T^c}^n|V_{T^c}^{0}}\prod_{k=1}^n p_{X_{T,k}|U_{T^c}^{k-1}, X_{T^c,k} } \right.\right) \notag\\
&=D_\lambda\left(p_{U_{T^c}^{0}}p_{\hat W_{T\times \{d\}}| V_{T^c}^{0},Y_{T^c}^n}\prod_{k=1}^n p_{X_{\mathcal{I},k},Y_{T^c,k}|U_{T^c}^{k-1}}\left\|p_{U_{T^c}^{0}}p_{\hat W_{T\times \{d\}}| V_{T^c}^{0},Y_{T^c}^n} \prod_{k=1}^n (s_{X_{T^c,k},Y_{T^c,k}|V_{T^c}^{k-1}} p_{X_{T,k}|U_{T^c}^{k-1}, X_{T^c,k}}) \right.\right) \notag\\
&\stackrel{\eqref{memorylessStatement}}{=} D_\lambda\left(p_{U_{T^c}^{0}}p_{\hat W_{T\times \{d\}}| V_{T^c}^{0},Y_{T^c}^n}\prod_{k=1}^n (p_{X_{\mathcal{I},k}|U_{T^c}^{k-1}} q_{Y_{T^c}|X_{\mathcal{I}}})\right\| \notag\\
&\qquad \quad \left. p_{U_{T^c}^{0}}p_{\hat W_{T\times \{d\}}| V_{T^c}^{0},Y_{T^c}^n} \prod_{k=1}^n (s_{X_{T^c,k}|V_{T^c}^{k-1}}   s_{Y_{T^c,k}|V_{T^c}^{k-1},X_{T^c,k}} p_{X_{T,k}|U_{T^c}^{k-1}, X_{T^c,k} })\right)\notag\\
&\stackrel{\text{(a)}}{=} D_\lambda\left(p_{U_{T^c}^{0}}p_{\hat W_{T\times \{d\}}| V_{T^c}^{0},Y_{T^c}^n}\prod_{k=1}^n (p_{X_{\mathcal{I},k}|U_{T^c}^{k-1}} q_{Y_{T^c}|X_{\mathcal{I}}})\right\| \notag\\
&\qquad \quad \left. p_{U_{T^c}^{0}}p_{\hat W_{T\times \{d\}}| V_{T^c}^{0},Y_{T^c}^n} \prod_{k=1}^n (s_{X_{T^c,k}|V_{T^c}^{k-1}}   s_{Y_{T^c,k}|X_{T^c,k}} p_{X_{T,k}|U_{T^c}^{k-1}, X_{T^c,k} })\right)\notag\\
&\stackrel{\text{(b)}}{=} D_\lambda\left(p_{U_{T^c}^{0}}p_{\hat W_{T\times \{d\}}| V_{T^c}^{0},Y_{T^c}^n}\prod_{k=1}^n (p_{X_{\mathcal{I},k}|U_{T^c}^{k-1}} q_{Y_{T^c}|X_{\mathcal{I}}})\right\| \notag\\
&\qquad \quad \left. p_{U_{T^c}^{0}}p_{\hat W_{T\times \{d\}}| V_{T^c}^{0},Y_{T^c}^n} \prod_{k=1}^n (p_{X_{T^c,k}|U_{T^c}^{k-1}} p_{X_{T,k}|U_{T^c}^{k-1}, X_{T^c,k}}   s_{Y_{T^c,k}| X_{T^c,k}})\right)\notag\\
& = D_\lambda\left(p_{U_{T^c}^{0}}p_{\hat W_{T\times \{d\}}| V_{T^c}^{0},Y_{T^c}^n}\prod_{k=1}^n (p_{X_{\mathcal{I},k}|U_{T^c}^{k-1}} q_{Y_{T^c}|X_{\mathcal{I}}})\right\| \notag\\
&\qquad \quad \left. p_{U_{T^c}^{0}}p_{\hat W_{T\times \{d\}}| V_{T^c}^{0},Y_{T^c}^n} \prod_{k=1}^n (p_{X_{\mathcal{I},k}|U_{T^c}^{k-1}}s_{Y_{T^c,k}| X_{T^c,k}})\right), \label{eqnDlambdaFirstChain}
 \end{align}
 where
 \begin{enumerate}
  \item[(a)] follows from Property~(iii) in Lemma~\ref{lemmaSimulatingDistribution}.
 \item[(b)] follows from Property~(v) in Lemma~\ref{lemmaSimulatingDistribution}.
 \end{enumerate}
 \subsection{Single-Letterizing the R\'{e}nyi Divergence}
Consider the distribution
\begin{align}
p_{U_{T^c}^0,X_{\mathcal{I}}^n, Y_{T^c}^n}& = p_{U_{T^c}^0}p_{X_{\mathcal{I}}^n, Y_{T^c}^n|U_{T^c}^0} \notag\\
& = p_{U_{T^c}^0} \prod_{k=1}^n p_{X_{\mathcal{I},k}, Y_{T^c,k}|U_{T^c}^{k-1}}\notag\\
& =  p_{U_{T^c}^0} \prod_{k=1}^n (p_{X_{\mathcal{I},k}|U_{T^c}^{k-1}}p_{ Y_{T^c,k}|U_{T^c}^{k-1},X_{\mathcal{I},k}})\notag\\
&\stackrel{\eqref{memorylessStatement}}{=}  p_{U_{T^c}^0} \prod_{k=1}^n (p_{X_{\mathcal{I},k}|U_{T^c}^{k-1}}q_{Y_{T^c}|X_{\mathcal{I}}}). \label{distribution1TempInProof}
\end{align}
Using \eqref{eqnDlambdaBeforeFirstChain}, \eqref{eqnDlambdaFirstChain} and Definition~\ref{defRenyiDivergence} and omitting subscripts of probability distributions to simplify notation, we have
\begin{align}
& D_\lambda(p_{W_{T},\hat W_{T\times \{d\}}}\|p_{W_{T}} s_{\hat W_{T\times \{d\}}})   \notag\\
&\le \frac{1}{\lambda-1}\log\left(\sum_{u_{T^c}^0,x_{\mathcal{I}}^n, y_{T^c}^n, \hat w_{T\times \{d\}}} \hspace{-0.4 in} p(u_{T^c}^{0})p(\hat w_{T\times \{d\}}| v_{T^c}^{0},y_{T^c}^n)\prod_{k=1}^n \left(p(x_{\mathcal{I},k}|u_{T^c}^{k-1}) q(y_{T^c,k}|x_{\mathcal{I},k})\left(\frac{q(y_{T^c,k}| x_{\mathcal{I},k})}{s(y_{T^c,k}| x_{T^c,k})}\right)^{\lambda-1}\right)\right) \notag\\
&\stackrel{\eqref{distribution1TempInProof}}{=}\frac{1}{\lambda-1}\log\left(\sum_{u_{T^c}^0,x_{\mathcal{I}}^n, y_{T^c}^n, \hat w_{T\times \{d\}}}p(u_{T^c}^0,x_{\mathcal{I}}^n, y_{T^c}^n)p(\hat w_{T\times \{d\}}| v_{T^c}^{0},y_{T^c}^n)\prod_{k=1}^n \left(\frac{q(y_{T^c,k}| x_{\mathcal{I},k})}{s(y_{T^c,k}| x_{T^c,k})}\right)^{\lambda-1}\right) \notag\\
&= \frac{1}{\lambda-1}\log\left(\sum_{x_{\mathcal{I}}^n, y_{T^c}^n}p(x_{\mathcal{I}}^n, y_{T^c}^n)\prod_{k=1}^n \left(\frac{q(y_{T^c,k}| x_{\mathcal{I},k})}{s(y_{T^c,k}| x_{T^c,k})}\right)^{\lambda-1}\right).  \label{eqnDlambdaSecondChain}
\end{align}
Following \eqref{eqnDlambdaSecondChain}, we consider the following chain of equalities:
\begin{align}
& \sum_{x_{\mathcal{I}}^n, y_{T^c}^n}p(x_{\mathcal{I}}^n, y_{T^c}^n)\prod_{k=1}^n \left(\frac{q(y_{T^c,k}| x_{\mathcal{I},k})}{s(y_{T^c,k}| x_{T^c,k})}\right)^{\lambda-1}\notag\\
& = \sum_{x_{\mathcal{I}}^n, y_{T^c}^n}\prod_{k=1}^n \left(p(x_{\mathcal{I},k},y_{T^c,k}| x_{\mathcal{I}}^{k-1}, y_{T^c}^{k-1}) \left(\frac{q(y_{T^c,k}| x_{\mathcal{I},k})}{s(y_{T^c,k}| x_{T^c,k})}\right)^{\lambda-1}\right)\notag\\
&\stackrel{\eqref{memorylessStatement}}{=} \sum_{x_{\mathcal{I}}^n, y_{T^c}^n}\prod_{k=1}^n \left(p(x_{\mathcal{I},k}| x_{\mathcal{I}}^{k-1}, y_{T^c}^{k-1}) \frac{\left(q(y_{T^c,k}| x_{\mathcal{I},k})\right)^\lambda}{\left(s(y_{T^c,k}| x_{T^c,k})\right)^{\lambda-1}}\right) \:. \label{eqnDlambdaBefore4thChain}
\end{align}
Letting $f_0^{(\lambda, T)}(x_{\mathcal{I}}^0, y_{T^c}^0)\triangleq 1$ and
\begin{equation}
f_k^{(\lambda, T)}(x_{\mathcal{I}}^k, y_{T^c}^k)\triangleq \prod_{\ell=1}^{k} \left(p(x_{\mathcal{I},\ell}| x_{\mathcal{I}}^{\ell-1}, y_{T^c}^{\ell-1}) \frac{\left(q(y_{T^c,\ell}| x_{\mathcal{I},\ell})\right)^\lambda}{\left(s(y_{T^c,\ell}| x_{T^c,\ell})\right)^{\lambda-1}}\right) \label{defFlambda}
\end{equation}
for each $k\in\{1, 2, \ldots, n\}$ and following \eqref{eqnDlambdaBefore4thChain}, we consider
\begin{align}
&\log\sum_{x_{\mathcal{I}}^n, y_{T^c}^n}\prod_{k=1}^n \left(p(x_{\mathcal{I},k}| x_{\mathcal{I}}^{k-1}, y_{T^c}^{k-1}) \frac{\left(q(y_{T^c,k}| x_{\mathcal{I},k})\right)^\lambda}{\left(s(y_{T^c,k}| x_{T^c,k})\right)^{\lambda-1}}\right) \notag\\
&
\stackrel{\eqref{defFlambda}}{=}\log\sum_{x_{\mathcal{I}}^n, y_{T^c}^n}f_n^{(\lambda, T)}(x_{\mathcal{I}}^n, y_{T^c}^n) \notag\\
& \stackrel{\text{(a)}}{=}  \log \prod_{k=1}^n \frac{\sum_{x_{\mathcal{I}}^k, y_{T^c}^k}f_k^{(\lambda, T)}(x_{\mathcal{I}}^k, y_{T^c}^k)}{\sum_{x_{\mathcal{I}}^{k-1}, y_{T^c}^{k-1}}f_{k-1}^{(\lambda, T)}(x_{\mathcal{I}}^{k-1}, y_{T^c}^{k-1})} \notag\\
&= \sum_{k=1}^n \log\left(\frac{\sum_{x_{\mathcal{I}}^k, y_{T^c}^k}f_k^{(\lambda, T)}(x_{\mathcal{I}}^k, y_{T^c}^k)}{\sum_{x_{\mathcal{I}}^{k-1}, y_{T^c}^{k-1}}f_{k-1}^{(\lambda, T)}(x_{\mathcal{I}}^{k-1}, y_{T^c}^{k-1})} \right), \label{eqnDlambda4thChain}
\end{align}
where (a) is a telescoping product.
For each $k\in\{1, 2, \ldots, n\}$, define $p_{X_{\mathcal{I},k}}^{(\lambda, T)}$ to be the following distribution:
\begin{align}
 p_{X_{\mathcal{I},k}}^{(\lambda, T)}(x_{\mathcal{I},k}) \triangleq \frac{  \sum_{x_{\mathcal{I}}^{k-1}, y_{T^c}^{k-1}} p(x_{\mathcal{I},k}|  x_{\mathcal{I}}^{k-1},   y_{T^c}^{k-1})f_{k-1}^{(\lambda, T)}(x_{\mathcal{I}}^{k-1}, y_{T^c}^{k-1})}{\sum_{x_{\mathcal{I}}^{k-1}, y_{T^c}^{k-1}}f_{k-1}^{(\lambda, T)}(x_{\mathcal{I}}^{k-1}, y_{T^c}^{k-1})}  \label{defPLambdaInProof}
\end{align}
for all $x_{\mathcal{I},k}$.
 Combining \eqref{eqnDlambdaBefore4thChain}, \eqref{eqnDlambda4thChain} and \eqref{defPLambdaInProof}, we obtain
\begin{align*}
\log\left(\sum_{x_{\mathcal{I}}^n, y_{T^c}^n}p(x_{\mathcal{I}}^n, y_{T^c}^n)\prod_{k=1}^n \left(\frac{q(y_{T^c,k}| x_{\mathcal{I},k})}{s(y_{T^c,k}| x_{T^c,k})}\right)^{\lambda-1}\right)  = \sum_{k=1}^n \log\left(\sum_{x_{\mathcal{I},k}}p_{X_{\mathcal{I},k}}^{(\lambda, T)}(x_{\mathcal{I},k})\sum_{y_{T^c,k}} \frac{\left(q(y_{T^c,k}| x_{\mathcal{I},k})\right)^\lambda}{\left(s(y_{T^c,k}| x_{T^c,k})\right)^{\lambda-1}} \right),
\end{align*}
which implies from \eqref{eqnDlambdaSecondChain} and Definition~\ref{defRenyiDivergence} that
\begin{equation}
 D_\lambda(p_{W_{T}, \hat W_{T\times \{d\}}}\|p_{W_{T}}s_{\hat W_{T\times \{d\}}}) \le \sum_{k=1}^n D_\lambda( q_{Y_{T^c}|X_{\mathcal{I}}}\|s_{Y_{T^c,k}|X_{T^c,k}}|p_{X_{\mathcal{I},k}}^{(\lambda, T)}). \label{eqnDLambdaFourthchain*}
\end{equation}
\subsection{Representing Distributions in the R\'enyi Divergence by a Single Distribution}
Construct a probability distribution $p_{X_{\mathcal{I},k}, Y_{T^c,k}}^{(\lambda, T)}$ for each $k\in\{1, 2, \ldots, n\}$ as
\begin{equation}
p_{X_{\mathcal{I},k}, Y_{T^c,k}}^{(\lambda, T)}(x_\mathcal{I}, y_{T^c})\triangleq p_{X_{\mathcal{I},k}}^{(\lambda, T)} (x_\mathcal{I})q_{Y_{T^c}|X_{\mathcal{I}}}(y_{T^c}|x_{\mathcal{I}}) \label{qLambdaDefining}
\end{equation}
for all $(x_\mathcal{I},y_{T^c})$ (cf.\ \eqref{defPLambdaInProof}), where $q_{Y_{T^c}|X_{\mathcal{I}}}$ denotes the channel of the DM-MMN.  Combining \eqref{defFlambda}, \eqref{defPLambdaInProof}, \eqref{qLambdaDefining} and Property~(iv) in Lemma~\ref{lemmaSimulatingDistribution}, we have
\begin{equation}
p_{X_{\mathcal{I},k},Y_{T^c,k}}^{(\lambda, T)} = s_{X_{\mathcal{I},k},Y_{T^c,k}}^{(\lambda, T)} \label{defPLambdaEqualSLambda}
\end{equation}
for each $k\in\{1, 2, \ldots, n\}$ where $s_{X_{\mathcal{I},k},Y_{T^c,k}}^{(\lambda, T)}$ is as defined in \eqref{defsRecursive**}. Then, it follows from Property~(iv) in Lemma~\ref{lemmaSimulatingDistribution}, \eqref{defPLambdaEqualSLambda} and \eqref{eqnDLambdaFourthchain*} that
\begin{equation}
D_\lambda(p_{W_{T}, \hat W_{T\times \{d\}}}\|p_{W_{T}}s_{\hat W_{T\times \{d\}}}) \le \sum_{k=1}^n D_\lambda( q_{Y_{T^c}|X_{\mathcal{I}}}\|p_{Y_{T^c,k}|X_{T^c,k}}^{(\lambda, T)}|p_{X_{\mathcal{I},k}}^{(\lambda, T)}).  \label{eqnDLambda5thchain}
\end{equation}
Using \eqref{qLambdaDefining} and Proposition~\ref{propositionMCsimplification}, we obtain
 \begin{equation}
p_{Y_{T^c,k}|X_{\mathcal{I},k}}^{(\lambda, T)}=q_{Y_{T^c}|X_{X_\mathcal{I}}}
 \label{qLambdaStatement2}
 \end{equation}
 for all $k\in\{1, 2, \ldots, n\}$, which implies from \eqref{eqnDLambda5thchain} that
\begin{equation}
 D_\lambda(p_{W_{T}, \hat W_{T\times \{d\}}}\|p_{W_{T}}s_{\hat W_{T\times \{d\}}}) \le \sum_{k=1}^n D_\lambda( p_{Y_{T^c,k}|X_{\mathcal{I},k}}^{(\lambda, T)}\|p_{Y_{T^c,k}|X_{T^c,k}}^{(\lambda, T)}|p_{X_{\mathcal{I},k}}^{(\lambda, T)}).  \label{eqnDLambdaFourthchain}
\end{equation}
\subsection{Introduction of a Time-sharing Random Variable}
Let $Q_n$ be a random variable uniformly distributed on $\{1, 2, \ldots, n\}$ and independent of all other random variables. Construct the probability distribution $p_{Q_n, X_\mathcal{I}^n, Y_{T^c}^n}^{(\lambda, T)}$ such that
\begin{equation}
p_{Q_n, X_\mathcal{I}^n, Y_{T^c}^n}^{(\lambda, T)}(k,x_{\mathcal{I}}^n, y_{T^c}^n)=\frac{1}{n}\prod_{h=1}^n p_{X_{\mathcal{I},h}, Y_{T^c,h}}^{(\lambda, T)}(x_{\mathcal{I},h}, y_{T^c,h}) \label{timeSharingVariable1}
\end{equation}
for all $k\in\{1, 2, \ldots, n\}$, $x_{\mathcal{I}}^n \in \mathcal{X}_{\mathcal{I}}^n$ and $y_{T^c}^n \in \mathcal{Y}_{T^c}^n$. Then, we can calculate the joint distributions $p_{Q_n, X_{\mathcal{I},Q_n}, Y_{T^c, Q_n}}^{(\lambda, T)}$ and $p_{X_{\mathcal{I},Q_n}, Y_{T^c, Q_n}}^{(\lambda, T)}$ as follows:
\begin{align}
p_{Q_n, X_{\mathcal{I},Q_n}, Y_{T^c, Q_n}}^{(\lambda, T)}(k,x_{\mathcal{I}}, y_{T^c}) & = p_{Q_n}^{(\lambda, T)}(k)p_{ X_{\mathcal{I},Q_n}, Y_{T^c, Q_n}|Q_n}^{(\lambda, T)}(x_{\mathcal{I}}, y_{T^c}|k)\notag\\
& = p_{Q_n}^{(\lambda, T)}(k)p_{ X_{\mathcal{I},k}, Y_{T^c, k}|Q_n}^{(\lambda, T)}(x_{\mathcal{I}}, y_{T^c}|k)\notag\\
&\stackrel{\eqref{timeSharingVariable1}}{=}  \frac{1}{n}p_{ X_{\mathcal{I},k}, Y_{T^c, k}}^{(\lambda, T)}(x_{\mathcal{I}}, y_{T^c}) \label{jointDistributionQnXY}
\end{align}
and
\begin{align}
p_{X_{\mathcal{I},Q_n}, Y_{T^c, Q_n}}^{(\lambda, T)}(x_{\mathcal{I}}, y_{T^c})  &\stackrel{\eqref{jointDistributionQnXY}}{=} \frac{1}{n}\sum_{k=1}^n p_{ X_{\mathcal{I},k}, Y_{T^c, k}}^{(\lambda, T)}(x_{\mathcal{I}}, y_{T^c}) \notag\\
& \stackrel{\eqref{qLambdaStatement2}}{=} \frac{1}{n}\sum_{k=1}^n p_{ X_{\mathcal{I},k}}^{(\lambda, T)}(x_{\mathcal{I}})q_{ Y_{T^c}|X_\mathcal{I}}(y_{T^c}|x_{\mathcal{I}})\notag\\
&\stackrel{\eqref{jointDistributionQnXY}}{=} p_{ X_{\mathcal{I},Q_n}}^{(\lambda, T)}(x_{\mathcal{I}}) q_{ Y_{T^c}|X_\mathcal{I}}(y_{T^c}|x_{\mathcal{I}}).  \label{jointDistributionQnXY*}
\end{align}
It follows from \eqref{jointDistributionQnXY} and Proposition~\ref{propositionMCsimplification} that
\begin{equation}
p_{X_{\mathcal{I},Q_n}, Y_{T^c, Q_n}|Q_n}^{(\lambda, T)}(x_{\mathcal{I}}, y_{T^c}|k)  = p_{ X_{\mathcal{I},k}, Y_{T^c, k}}^{(\lambda, T)}(x_{\mathcal{I}}, y_{T^c}), \label{jointDistributionQnXY***}
\end{equation}
\begin{equation}
\left(Q_n\rightarrow X_{\mathcal{I}, Q_n}\rightarrow Y_{T^c, Q_n}\right)_{s^{(\lambda, T)}} \label{markovchainProofStatement1}
\end{equation}
and
\begin{equation}
\left(Q_n\rightarrow X_{T^c, Q_n}\rightarrow Y_{T^c, Q_n}\right)_{s^{(\lambda, T)}}. \label{markovchainProofStatement1*}
\end{equation}
 Following \eqref{eqnDLambdaFourthchain}, consider the following chain of inequalities:
\begin{align}
&\frac{1}{n}\sum_{k=1}^n D_\lambda( p_{Y_{T^c,k}|X_{\mathcal{I},k}}^{(\lambda, T)}\|p_{Y_{T^c,k}|X_{T^c,k}}^{(\lambda, T)}|p_{X_{\mathcal{I},k}}^{(\lambda, T)}) \notag\\
& = \frac{1}{(\lambda-1)n}\sum_{k=1}^n \log \sum_{x_{\mathcal{I}}} p_{X_{\mathcal{I},k}}^{(\lambda, T)}(x_\mathcal{I}) \sum_{y_{T^c}} \frac{(p_{Y_{T^c,k}|X_{\mathcal{I},k}}^{(\lambda, T)}(y_{T^c}|x_{\mathcal{I}}))^\lambda}{(p_{Y_{T^c,k}|X_{T^c,k}}^{(\lambda, T)}(y_{T^c}|x_{T^c}))^{\lambda-1}} \notag\\
& \stackrel{\eqref{jointDistributionQnXY***}}{=} \frac{1}{(\lambda-1)n}\sum_{k=1}^n \log \sum_{x_{\mathcal{I}}} p_{X_{\mathcal{I},Q_n}|Q_n}^{(\lambda, T)}(x_\mathcal{I}|k)   \sum_{y_{T^c}} \frac{(p_{Y_{T^c,Q_n}|X_{\mathcal{I},Q_n},Q_n}^{(\lambda, T)}(y_{T^c}|x_{\mathcal{I}}, k))^\lambda}{(p_{Y_{T^c,Q_n}|X_{T^c,Q_n}, Q_n}^{(\lambda, T)}(y_{T^c}|x_{T^c}, k))^{\lambda-1}} \notag\\
&\stackrel{\text{(a)}}{\le } \frac{1}{(\lambda-1)} \log \sum_{k=1}^n \frac{1}{n} \sum_{x_{\mathcal{I}}} p_{X_{\mathcal{I},Q_n}|Q_n}^{(\lambda, T)}(x_\mathcal{I}|k)   \sum_{y_{T^c}} \frac{(p_{Y_{T^c,Q_n}|X_{\mathcal{I},Q_n},Q_n}^{(\lambda, T)}(y_{T^c}|x_{\mathcal{I}}, k))^\lambda}{(p_{Y_{T^c,Q_n}|X_{T^c,Q_n}, Q_n}^{(\lambda, T)}(y_{T^c}|x_{T^c}, k))^{\lambda-1}} \notag\\
&\stackrel{\text{(b)}}{=} \frac{1}{(\lambda-1)} \log \sum_{k=1}^n \sum_{x_{\mathcal{I}}} p_{X_{\mathcal{I},Q_n},Q_n}^{(\lambda, T)}(x_\mathcal{I},k)   \sum_{y_{T^c}} \frac{(p_{Y_{T^c,Q_n}|X_{\mathcal{I},Q_n},Q_n}^{(\lambda, T)}(y_{T^c}|x_{\mathcal{I}}, k))^\lambda}{(p_{Y_{T^c,Q_n}|X_{T^c,Q_n}, Q_n}^{(\lambda, T)}(y_{T^c}|x_{T^c}, k))^{\lambda-1}} \notag\\
&\stackrel{\text{(c)}}{=} \frac{1}{(\lambda-1)} \log \sum_{k=1}^n \sum_{x_{\mathcal{I}}} p_{X_{\mathcal{I},Q_n},Q_n}^{(\lambda, T)}(x_\mathcal{I},k)   \sum_{y_{T^c}} \frac{(p_{Y_{T^c,Q_n}|X_{\mathcal{I},Q_n}}^{(\lambda, T)}(y_{T^c}|x_{\mathcal{I}}))^\lambda}{(p_{Y_{T^c,Q_n}|X_{T^c,Q_n}}^{(\lambda, T)}(y_{T^c}|x_{T^c}))^{\lambda-1}} \notag\\
&=\frac{1}{(\lambda-1)} \log \sum_{x_{\mathcal{I}}} p_{X_{\mathcal{I},Q_n}}^{(\lambda, T)}(x_\mathcal{I})   \sum_{y_{T^c}} \frac{(p_{Y_{T^c,Q_n}|X_{\mathcal{I},Q_n}}^{(\lambda, T)}(y_{T^c}|x_{\mathcal{I}}))^\lambda}{(p_{Y_{T^c,Q_n}|X_{T^c,Q_n}}^{(\lambda, T)}(y_{T^c}|x_{T^c}))^{\lambda-1}} \notag\\
& = D_\lambda( p_{Y_{T^c, Q_n}|X_{\mathcal{I}, Q_n}}^{(\lambda, T)}\|p_{Y_{T^c,Q_n}|X_{T^c, Q_n}}^{(\lambda, T)}|p_{X_{\mathcal{I},Q_n}}^{(\lambda, T)}), \label{eqnDLambdaFifthchain}
\end{align}
where
\begin{enumerate}
\item[(a)] follows from the concavity of $t\mapsto \log t$ and Jensen's inequality.
\item[(b)] follows from \eqref{timeSharingVariable1} that $p_{Q_n}^{(\lambda, T)}(k)=1/n$ for all $k\in\{1, 2, \ldots, n\}$.
    \item[(c)] follows from \eqref{markovchainProofStatement1} and \eqref{markovchainProofStatement1*}.
\end{enumerate}
Combining \eqref{eqnDLambdaReversechain}, \eqref{eqnDLambdaFourthchain} and \eqref{eqnDLambdaFifthchain}, we obtain
\begin{equation}
\sum_{i\in T}nR_i + \lambda(\lambda-1)^{-1}\log(1-\bar \epsilon) \le n D_\lambda( p_{Y_{T^c, Q_n}|X_{\mathcal{I}, Q_n}}^{(\lambda, T)}\|p_{Y_{T^c,Q_n}|X_{T^c, Q_n}}^{(\lambda, T)}|p_{X_{\mathcal{I},Q_n}}^{(\lambda, T)}) \label{eqnDLambdaSixthchain}
\end{equation}
for all $\lambda\in (1, \infty)$.
\subsection{Approximating the R\'enyi Divergence by Conditional Relative Entropy}
For each block length $n$, choose $\lambda$ to be dependent on $n$ as follows:
\begin{equation*}
\lambda_n \triangleq 1+\frac{1}{\sqrt{n}}. 
\end{equation*}
It then follows from \eqref{eqnDLambdaSixthchain}, Proposition~\ref{propositionDlambdaToMutualInfo}, and the fact that $|\mathcal{X}_{T} ||\mathcal{Y}_{T^c}|\le|\mathcal{X}_{\mathcal{I}} ||\mathcal{Y}_{\mathcal{I}}|$  that
\begin{align}
\sum_{i\in T}R_i + \left(\frac{1}{n}+\frac{1}{\sqrt{n}}\right)\log(1-\bar \epsilon) &\le   D_{\lambda_n} ( p_{Y_{T^c, Q_n}|X_{\mathcal{I}, Q_n}}^{(\lambda_n, T)}\|p_{Y_{T^c,Q_n}|X_{T^c, Q_n}}^{(\lambda_n, T)}|p_{X_{\mathcal{I},Q_n}}^{(\lambda_n, T)}) \nonumber\\
&\le D( p_{Y_{T^c, Q_n}|X_{\mathcal{I}, Q_n}}^{(\lambda_n, T)}\|p_{Y_{T^c,Q_n}|X_{T^c, Q_n}}^{(\lambda_n, T)}|p_{X_{\mathcal{I},Q_n}}^{(\lambda_n, T)}) + \frac{8(|\mathcal{X}_{\mathcal{I}} ||\mathcal{Y}_{\mathcal{I}} |)^5}{\sqrt{n}} \label{eqnDLambda7thchain}
\end{align}
if $n\ge 16$ (i.e., $\lambda_n \le 5/4 $ so Proposition~\ref{propositionDlambdaToMutualInfo} applies). Taking the limit inferior on both sides of \eqref{eqnDLambda7thchain}, we obtain
\begin{equation}
\sum_{i\in T}R_i \le \liminf_{n\rightarrow \infty} D(p_{Y_{T^c, Q_n}|X_{\mathcal{I}, Q_n}}^{(\lambda_n, T)}\|p_{Y_{T^c,Q_n}|X_{T^c, Q_n}}^{(\lambda_n, T)}|p_{X_{\mathcal{I},Q_n}}^{(\lambda_n, T)}). \label{eqnDLambda8thchain}
\end{equation}
Consider each distribution on $(\mathcal{X}_{\mathcal{I}}, \mathcal{Y}_{T^c})$ as a point in the $|\mathcal{X}_{\mathcal{I}}||\mathcal{Y}_{T^c}|$-dimensional Euclidean space. Then, by the compactness of the probability simplex, there exists a subsequence of the natural numbers $\{1, 2, \ldots\}$, say indexed by  $\{n_\ell\}_{\ell=1}^\infty$, such that $\{p_{X_{\mathcal{I},Q_{n_\ell}},Y_{T^c,Q_{n_\ell}}}^{(\lambda_{n_\ell}, T)}\}_{\ell=1}^\infty$ is convergent with respect to the $\mathcal{L}_1$-distance. Let $\bar p_{X_{\mathcal{I}},Y_{T^c}}$ be the limit of the subsequence such that
\begin{equation}
\bar p_{X_{\mathcal{I}},Y_{T^c}}(x_\mathcal{I},y_{T^c}) =\lim_{\ell \rightarrow \infty} p_{X_{\mathcal{I},Q_{n_\ell}},Y_{T^c,Q_{n_\ell}}}^{(\lambda_{n_\ell}, T)}(x_\mathcal{I},y_{T^c}) \label{jointDistributionBars}
\end{equation}
for all $(x_\mathcal{I},y_{T^c})$. Combining \eqref{jointDistributionQnXY*} and \eqref{jointDistributionBars}, we have
\begin{equation}
\bar p_{X_{\mathcal{I}},Y_{T^c}}(x_\mathcal{I},y_{T^c}) = \bar p_{X_{\mathcal{I}}}(x_\mathcal{I}) q_{Y_{T^c}|X_{\mathcal{I}}}(y_{T^c}|x_{\mathcal{I}}). \label{thmMainResultfinalStatement1}
\end{equation}
Since $D(p_{Y_{T^c}|X_{\mathcal{I}}}\|p_{Y_{T^c}|X_{T^c}}|p_{X_{\mathcal{I}}})$ is a continuous functional of distribution $p_{X_\mathcal{I}, Y_{T^c}}$, it follows from \eqref{eqnDLambda8thchain} and \eqref{jointDistributionBars} that
\begin{align}
\sum_{i\in T}R_i & \le D(\bar p_{Y_{T^c}|X_{\mathcal{I}}}\|\bar p_{Y_{T^c}|X_{T^c}}|\bar p_{X_{\mathcal{I}}}) \notag\\
&= \sum_{x_\mathcal{I}} \bar  p_{X_{\mathcal{I}}}(x_\mathcal{I})\sum_{y_{T^c}}\bar p_{Y_{T^c}|X_{\mathcal{I}}}\log\frac{\bar p_{Y_{T^c}|X_{\mathcal{I}}}(y_{T^c}|x_\mathcal{I})}{\bar p_{Y_{T^c}|X_{T^c}}(y_{T^c}|x_{T^c})} \notag\\
&= \sum_{x_\mathcal{I},y_{T^c}}\bar p_{X_{\mathcal{I}},Y_{T^c}}(x_{\mathcal{I}},y_{T^c})\log\frac{\bar p_{X_{T},Y_{T^c}|X_{T^c}}(x_{T},y_{T^c}|x_{T^c})}{\bar p_{X_T|X_{T^c}}(x_T|x_{T^c})\bar p_{Y_{T^c}|X_{T^c}}(y_{T^c}|x_{T^c})} \notag\\
& = I_{\bar p_{X_{\mathcal{I}}, Y_{T^c}}}(X_T; Y_{T^c}|X_{T^c}). \label{thmMainResultfinalStatement2}
\end{align}
The theorem then follows from \eqref{thmMainResultfinalStatement1} and \eqref{thmMainResultfinalStatement2}.

\subsection{Comparison to the Proof of the Cut-Set Bound using Fano's Inequality} \label{sec:fano}
Following the setting in Section~\ref{subsectionLowerBound} at the beginning of the proof of Theorem~\ref{thmMainResult} and following the cut-set bound approach that uses Fano's inequality \cite[Theorem 18.1]{elgamal} (leading to a weak converse), we can lower bound the average error probability $\epsilon_n$ as follows
\begin{equation}
(1 - \epsilon_n) \sum_{i\in T}n R_i \le I_{p_{W_\mathcal{I}, Y_\mathcal{I}^n}}(W_T; Y_{T^c}^n|W_{T^c}). \label{elucidateTemp1}
\end{equation}
The bound \eqref{elucidateTemp1} holds  for each $T$ that satisfies $T^c\cap \mathcal{D}\ne \emptyset$.
Next, using the DPI for the relative entropy and a time-sharing random variable for the purpose of  single-letterization \cite[Theorem 18.1]{elgamal}, it can be shown that
\begin{equation}
I_{p_{W_\mathcal{I}, Y_\mathcal{I}^n}}(W_T; Y_{T^c}^n|W_{T^c}) \le n I_{\bar p_{X_\mathcal{I}}q_{Y_\mathcal{I}|X_\mathcal{I}}}(X_{T}; Y_{T^c}|X_{T^c}), \label{elucidateTemp2}
\end{equation}
where $\bar p_{X_\mathcal{I}}(x_\mathcal{I})=\frac{1}{n}\sum_{k=1}^n p_{X_{\mathcal{I},k}}(x_\mathcal{I})$ is the empirical input distribution induced by the $(n, R_\mathcal{I})$-code. Combining \eqref{elucidateTemp1} and \eqref{elucidateTemp2} and using the fact that $\bar p_{X_\mathcal{I}}(x_\mathcal{I})$ does not depend on~$T$, we obtain
\begin{align}
\mathcal{C}_\epsilon \subseteq   \bigcup_{ \bar p_{X_{\mathcal{I}}}}  \bigcap_{T\subseteq \mathcal{I}: T^c \cap \mathcal{D} \ne \emptyset }\left\{ R_\mathcal{I}\left| \: \parbox[c]{2.6 in}{$ \sum_{ i\in T} R_{i}
 \le  \frac{1}{1-\epsilon}{I_{\bar p_{X_{\mathcal{I}}}q_{Y_{T^c}|X_\mathcal{I}}}(X_T; Y_{T^c}|X_{T^c})},\\
 R_i=0 \text{ for all }i\in\mathcal{S}^c$} \right.\right\}. \label{elucidateCutSet}
 \end{align}
For $\epsilon=0$, \eqref{elucidateCutSet} immediately reduces to the cut-set bound \eqref{cutSetBound}. For $\epsilon>0$, the bound in \eqref{elucidateCutSet} cannot be used to prove strong converse theorems because of the multiplicative factor $\frac{1}{1-\epsilon}$.

The proofs of~\eqref{strongConverseBound} and~\eqref{elucidateCutSet} share many common steps, but significantly they differ in the first step where for fixed rates, lower bounds on the error probabilities are sought. More specifically,  our approach relates a  conditional R\'{e}nyi divergence to the error probability (cf.\ Proposition~\ref{propositionDlambdaLowerBound}), while the approach that hinges  on Fano's inequality relates a conditional mutual information   to the error probability (cf.\ the inequality in~\eqref{elucidateTemp1}). However beyond the first  step, the application of the DPI and the method of single-letterization are almost the same for both proofs, but we do need to eventually approximate the conditional R\'enyi entropy with the conditional mutual information (cf.\ Proposition~\ref{propositionDlambdaToMutualInfo}) to obtain bound~\eqref{strongConverseBound}. The two different ways of lower bounding the error probability yield two different outer bounds stated in \eqref{strongConverseBound} and \eqref{elucidateCutSet} respectively.

\section{Classes of Multimessage Multicast Networks with Tight Cut-Set Bound}\label{sectionMMNwithTightCutSet}
In this section, we will use Theorem~\ref{thmMainResult} to prove strong converses for some classes of DM-MMNs whose capacity regions are known. Unless specified otherwise, we let $(\mathcal{S}, \mathcal{D})$ denote the multicast demand on the networks.
\subsection{Multicast Networks with Maximal Cut-Set Distribution} \label{sectionMulticastNetworks}
 We start this section by stating an achievability result for multimessage multicast networks in the following theorem, which is a specialization of the main result of {\em noisy network coding} by Lim, Kim, El Gamal and Chung~\cite{noisyNetworkCoding}. Noisy network coding was also discovered by Yassaee and Aref~\cite{NNCv2}.

\begin{Theorem} \label{theoremNNC}
 Let $(\mathcal{X}_\mathcal{I}, \mathcal{Y}_\mathcal{I}, q_{Y_\mathcal{I}|X_\mathcal{I}})$ be a DM-MMN, and let
\begin{equation}
\mathcal{R}_{\text{in}} \triangleq   \bigcup_{\prod_{i=1}^N p_{X_i}} \bigcap_{T\subseteq \mathcal{I}: T^c \cap \mathcal{D} \ne \emptyset }\left\{ R_\mathcal{I}\left| \: \parbox[c]{4 in}{$ \sum_{ i\in T}  R_{i}
 \le  I_{p_{X_{\mathcal{I}}}q_{Y_{T^c}|X_\mathcal{I}}}(X_T; Y_{T^c}|X_{T^c})-H_{p_{X_{\mathcal{I}}}q_{Y_{\mathcal{I}}|X_\mathcal{I}}}(Y_T|X_\mathcal{I}, Y_{T^c}) \linebreak \text{ where $p_{X_\mathcal{I}}\triangleq \prod_{i=1}^N p_{X_i}$,}\\
 R_i=0 \text{ for all }i\in\mathcal{S}^c$} \right.\right\}. \label{Rin}
\end{equation}
Then, $\mathcal{R}_{\text{in}} \subseteq \mathcal{C}_0$.
\end{Theorem}
\begin{IEEEproof}
The theorem follows by taking $\hat Y=Y$ in Theorem 1 of \cite{noisyNetworkCoding}.
\end{IEEEproof}

We would like to identify multicast networks whose inner bounds $\mathcal{R}_{\text{in}}$ coincides with our outer bound   $\mathcal{R}_{\text{out}}$ in Theorem~\ref{thmMainResult}. Using the following definition and corollary, we can state, in Theorem~\ref{thmMainResultApplication}, a sufficient condition for $\mathcal{R}_{\text{in}}=\mathcal{R}_{\text{out}}$ to hold.

\begin{Definition} \label{defMaximalProductDistribution}
A DM-MMN $(\mathcal{X}_\mathcal{I}, \mathcal{Y}_\mathcal{I}, q_{Y_{\mathcal{I}}|X_{\mathcal{I}}})$ is said to be \textit{dominated by a maximal product distribution} if there exists some product distribution $p_{X_\mathcal{I}}^*\triangleq \prod_{i=1}^N p_{X_i}^*$ such that the following statement holds for each $T\subseteq\mathcal{I}$:
\begin{align*}
 I_{p_{X_{\mathcal{I}}}^*q_{Y_{T^c}|X_\mathcal{I}}}(X_T; Y_{T^c}|X_{T^c})=  \max_{p_{X_\mathcal{I}}} \{ I_{p_{X_{\mathcal{I}}}q_{Y_{T^c}|X_\mathcal{I}}}(X_T; Y_{T^c}|X_{T^c})\}.
\end{align*}
\end{Definition}

The following corollary is a direct consequence of Theorem~\ref{thmMainResult} and Definition~\ref{defMaximalProductDistribution}, and the proof is deferred to Appendix~\ref{appendixB}.
\begin{Corollary} \label{corollaryProductDistributionOuterBound}
Let $(\mathcal{X}_\mathcal{I}, \mathcal{Y}_\mathcal{I}, q_{Y_{\mathcal{I}}|X_{\mathcal{I}}})$ be a DM-MMN, and let
\begin{equation}
\mathcal{R}_{\text{out}}^* \triangleq  \bigcup_{\prod_{i=1}^N p_{X_{i}}} \bigcap_{T\subseteq \mathcal{I}: T^c \cap \mathcal{D} \ne \emptyset }  \left\{ R_\mathcal{I}\left| \: \parbox[c]{2.7 in}{$ \sum_{ i\in T} R_{i}
 \le  I_{(\prod_{i=1}^N p_{X_{i}})q_{Y_{T^c}|X_\mathcal{I}}}(X_T; Y_{T^c}|X_{T^c}),\\
 R_i=0 \text{ for all }i\in\mathcal{S}^c$} \right.\right\}. \label{RoutDet}
\end{equation}
 If the DM-MMN is dominated by a maximal product distribution, then $\mathcal{C}_\epsilon \subseteq \mathcal{R}_{\text{out}}^* $ for all $\epsilon \in [0,1)$.
\end{Corollary}

\begin{Theorem} \label{thmMainResultApplication}
Let $(\mathcal{X}_\mathcal{I}, \mathcal{Y}_\mathcal{I}, q_{Y_{\mathcal{I}}|X_{\mathcal{I}}})$ be a DM-MMN. Suppose the DM-MMN satisfies the following two conditions: \begin{enumerate}
\item[1.] The DM-MMN is dominated by a maximal product distribution.
\item[2.] For all $T\subseteq \mathcal{I}$ and all $p_{X_{\mathcal{I}}}$,
$H_{p_{X_{\mathcal{I}}}q_{Y_{\mathcal{I}}|X_\mathcal{I}}}(Y_T|X_\mathcal{I}, Y_{T^c})=0$. 
 \end{enumerate}
 Then $\mathcal{R}_{\text{in}} =\mathcal{C}_\epsilon = \mathcal{R}_{\text{out}}^*$ for all $\epsilon \in [0,1)$.
\end{Theorem}
\begin{IEEEproof}
Since the DM-MMN is dominated by a maximal product distribution, it follows from Theorem~\ref{theoremNNC} and Corollary~\ref{corollaryProductDistributionOuterBound} that $\mathcal{R}_{\text{in}}\subseteq \mathcal{C}_0 \subseteq \mathcal{C}_\epsilon \subseteq \mathcal{R}_{\text{out}}^*$ for all $\epsilon \in [0,1)$. In addition, it follows from \eqref{Rin}, \eqref{RoutDet} and Condition 2 that $\mathcal{R}_{\text{in}}=\mathcal{R}_{\text{out}}^*$.
\end{IEEEproof}

Theorem~\ref{thmMainResultApplication} implies the strong converse for the classes of  DM-MMNs which satisfy Conditions 1 and 2. Since the deterministic relay networks with no interference \cite{multicastCapacityRelayNetworks}, the finite-field linear deterministic networks \cite{AvestimehrDeterministic,linearFiniteField09} and the wireless erasure networks \cite{dana06} satisfy both conditions in Theorem~\ref{thmMainResultApplication}, the strong converse holds for these networks. We note that for the class of wireless erasure networks, one  assumes that the erasure pattern of the entire network is known to each destination, i.e., $Y_{d}$ contains the erasure pattern as side information for each $d\in \mathcal{D}$ \cite[Section~III.C]{dana06}, and hence Condition 2 in Theorem~\ref{thmMainResultApplication} is satisfied. In the following subsection, we introduce a DM-MMN connected by independent DMCs and prove the strong converse using Corollary~\ref{corollaryProductDistributionOuterBound} and Theorem~\ref{thmMainResult}.

\subsection{DM-MMN Consisting of Independent DMCs} \label{sectionDM-MMNconsistingOfDMCs}
 Consider a DM-MMN where a DMC  is defined for every link $(i,j)\in \mathcal{I}\times \mathcal{I}$. Let $\mathcal{X}_{i,j}$ and $\mathcal{Y}_{i,j}$ denote the input and output alphabets of the DMC carrying information from node~$i$ to node~$j$ for each $(i,j)\in \mathcal{I}\times \mathcal{I}$, and let $q_{Y_{i,j}|X_{i,j}}$ denote the DMC. For each $(i,j)\in \mathcal{I}\times \mathcal{I}$, the capacity of channel $q_{Y_{i,j}|X_{i,j}}$, denoted by $C_{i,j}$, is attained by some $\bar p_{X_{i,j}}$, i.e.,
 \begin{align}
 C_{i,j}& \triangleq \max_{p_{X_{i,j}}}I_{p_{X_{i,j}} q_{Y_{i,j}|X_{i,j}}}(X_{i,j};Y_{i,j}) \
 \notag\\
 &= I_{ \bar p_{X_{i,j}} q_{Y_{i,j}|X_{i,j}}}(X_{i,j};Y_{i,j}). \label{defPTPCapacity}
 \end{align}
 Then, we define the input and output alphabets for each node $i$ in the following natural way:
\begin{equation*}
\mathcal{X}_i \triangleq \mathcal{X}_{i,1}\times \mathcal{X}_{i,2} \times \ldots \times \mathcal{X}_{i,N}
\end{equation*}
and
\begin{equation}
\mathcal{Y}_i \triangleq \mathcal{Y}_{1,i}\times \mathcal{X}_{2,i} \times \ldots \times \mathcal{X}_{N,i}\label{alphabetYSequence}
\end{equation}
for each $i\in\mathcal{I}$, and we let $q_{Y_\mathcal{I}|X_\mathcal{I}}$ denote the channel of the network. In addition, we assume
\begin{equation}
q_{Y_\mathcal{I}|X_\mathcal{I}} = \prod_{(i,j)\in\mathcal{I}\times \mathcal{I}}q_{Y_{i,j}|X_{i,j}}, \label{defPTPnetworkChannel}
\end{equation}
i.e., the random transformations (noises) from $X_{i,j}$ to $Y_{i,j}$  are independent and the overall channel of the network is in a product form. It then follows from \eqref{defPTPnetworkChannel} and Proposition~\ref{propositionMCsimplification} that
\begin{equation}
( (\{X_{k, \ell}\}_{(k, \ell)\ne (i,j)}, \{Y_{k, \ell}\}_{(k, \ell)\ne (i,j)} ) \rightarrow X_{i,j} \rightarrow Y_{i,j})_{p_{X_\mathcal{I}}q_{Y_\mathcal{I}|X_\mathcal{I}}} \label{markovChainPTPNetwork}
\end{equation}
forms a Markov chain for all $(i,j)\in \mathcal{I}\times \mathcal{I}$.
We call the network described above the \textit{DM-MMN consisting of independent DMCs}. One important example of such networks is the {\em line network} in which $\mathcal{I} \times \mathcal{I}$ consists of nonzero-capacity links of the form $(i,i+1)$ for all $i \in\{1,2,\ldots, N-1\}$ and zero-capacity links for the other node pairs. Define
\begin{equation}
\mathcal{R}^{\prime} \triangleq \bigcap_{T\subseteq \mathcal{I}: T^c \cap \mathcal{D} \ne \emptyset } \left\{ R_\mathcal{I}\left| \: \parbox[c]{1.8 in}{$ \sum_{ i\in T} R_{i}
 \le  \sum_{ (i,j)\in T \times T^c} C_{i,j},\\
 R_i=0 \text{ for all }i\in\mathcal{S}^c$} \right.\right\}. \label{RPrime}
\end{equation}
Since the DMCs from $X_{i,j}$ to $Y_{i,j}$ are all independent and each of the DMC can carry information at a rate arbitrarily close to the capacity, it follows from the network equivalence theory \cite{networkEquivalencePartI} and Theorem~\ref{theoremNNC} that $\mathcal{R}^{\prime}$ is precisely the capacity region of the DM-MMN consisting of independent DMCs, which is formally stated in the following corollary and proved in Appendix~\ref{appendixC}.

\begin{Corollary} \label{corollaryInnerBoundPTP}
$\mathcal{R}^{\prime}=\mathcal{C}_0$.
\end{Corollary}

We use the outer bound $\mathcal{R}_{\text{out}}$ proved in Theorem~\ref{thmMainResult} (cf.\ \eqref{Rout}) to prove the following lemma.

\begin{Lemma} \label{lemmaPTP}
$\mathcal{C}_\epsilon \subseteq \mathcal{R}_{\text{out}} \subseteq \mathcal{R}^{\prime}$ for all $\epsilon \in [0,1)$.
\end{Lemma}

For completeness, the  proof is provided in Appendix~\ref{appendixD}. The following theorem is a direct consequence of Corollary~\ref{corollaryInnerBoundPTP} and Lemma~\ref{lemmaPTP}.

\begin{Theorem} \label{theoremPTPNetwork}
Let $(\mathcal{X}_\mathcal{I}, \mathcal{Y}_\mathcal{I}, q_{Y_\mathcal{I}|X_\mathcal{I}})$ be a DM-MMN consisting of independent DMCs.
Then, $\mathcal{C}_\epsilon = \mathcal{R}^{\prime}$ for all $\epsilon \in [0,1)$.
\end{Theorem}

Theorem~\ref{theoremPTPNetwork} implies the strong converse for the class of DM-MMNs consisting of independent DMCs.
\subsection{Single-Destination DM-MMN Consisting of Independent DMCs with Destination Feedback} \label{sectionWithFeedback}
In this section, we examine a class of DM-MMNs with destination feedback, which is a generalization of the DM-MMN consisting of independent DMCs discussed in the previous section. We assume $|\mathcal{D}|=1$ and let $d \in \mathcal{I}$ denote the (single) destination node throughout this section. We define the single-destination DM-MMN consisting of independent DMCs with feedback as follows.

\begin{Definition} \label{defDM-MMNwithFeedback}
 Let $(\mathcal{X}_\mathcal{I}, \mathcal{Y}_\mathcal{I}, q_{Y_\mathcal{I}|X_\mathcal{I}})$ be DM-MMN consisting of independent DMCs with multicast demand $(\mathcal{S}, \{d\})$ as defined in the previous section. A single-destination DM-MMN with multicast demand $(\mathcal{S}, \{d\})$, denoted by $(\mathcal{X}_\mathcal{I}, \tilde {\mathcal{Y}}_\mathcal{I}, \tilde q_{\tilde Y_\mathcal{I}|X_\mathcal{I}})$, is called the \textit{feedback version of $(\mathcal{X}_\mathcal{I}, \mathcal{Y}_\mathcal{I}, q_{Y_\mathcal{I}|X_\mathcal{I}})$} if the following two conditions hold:
\begin{enumerate}
\item $\tilde {\mathcal{Y}}_i = \mathcal{Y}_i \times \mathcal{Y}_d$ for all $i\in\mathcal{I}$.
\item Suppose $(X_\mathcal{I}, Y_\mathcal{I})\in \mathcal{X}_\mathcal{I}\times \mathcal{Y}_\mathcal{I}$ associated with the MMN $(\mathcal{X}_\mathcal{I}, \mathcal{Y}_\mathcal{I}, q_{Y_\mathcal{I}|X_\mathcal{I}})$ is generated according to $p_{X_\mathcal{I}} q_{ Y_\mathcal{I}| X_\mathcal{I}}$ for some input distribution $p_{X_\mathcal{I}}$. Then, the random tuple $(\tilde X_\mathcal{I}, \tilde {Y}_\mathcal{I})\in \mathcal{X}_\mathcal{I}\times \tilde{\mathcal{Y}}_\mathcal{I}$ associated with the MMN $(\mathcal{X}_\mathcal{I}, \tilde {\mathcal{Y}}_\mathcal{I}, \tilde q_{\tilde Y_\mathcal{I}|X_\mathcal{I}})$ is distributed according to $p_{X_\mathcal{I}}\tilde q_{\tilde Y_\mathcal{I}|X_\mathcal{I}}$ where $\tilde X_i = X_i$ and $\tilde Y_i = (Y_i, Y_d)$ for all $i\in\mathcal{I}$.
\end{enumerate}
\end{Definition}

Let $(\mathcal{X}_\mathcal{I}, \tilde {\mathcal{Y}}_\mathcal{I}, \tilde q_{\tilde Y_\mathcal{I}|X_\mathcal{I}})$ be the feedback version of $(\mathcal{X}_\mathcal{I}, \mathcal{Y}_\mathcal{I}, q_{Y_\mathcal{I}|X_\mathcal{I}})$ with multicast demand $(\mathcal{S}, \{d\})$. It then follows from Definitions~\ref{defDM-MMNwithFeedback} and~\ref{defCode} that for any $(n, R_\mathcal{I})$-code on $(\mathcal{X}_\mathcal{I}, \tilde {\mathcal{Y}}_\mathcal{I}, \tilde q_{\tilde Y_\mathcal{I}|X_\mathcal{I}})$, both $Y_d^{k-1}$ and $Y_{i}^{k-1}$ are available for encoding $X_{i,k}$ at node~$i$ for all $i\in\mathcal{I}$. In other words, there exists for each $i\in\mathcal{I}$ a perfect feedback link which carries the output symbols at node~$d$ to node~$i$. Consequently, the capacity region of $(\mathcal{X}_\mathcal{I}, \mathcal{Y}_\mathcal{I}, q_{Y_\mathcal{I}|X_\mathcal{I}})$ is always a subset of the capacity region of $(\mathcal{X}_\mathcal{I}, \tilde {\mathcal{Y}}_\mathcal{I}, \tilde q_{\tilde Y_\mathcal{I}|X_\mathcal{I}})$. Shannon showed in \cite{Shannon56} that the capacity of any DMC is equal to the capacity of the feedback version, and the strong converse for the feedback version has been shown in \cite[Section~IV]{yuryRenyiDivergence}.  Also see \cite[Problem 2.5.16(c)]{Csi97} for another proof sketch of the strong converse for a DMC with feedback. Here, we show that $\mathcal{R}^{\prime}$ (defined in \eqref{RPrime}) is equal to the $\epsilon$-capacity region of any single-destination DM-MMN consisting of independent DMCs as well as the $\epsilon$-capacity region of the feedback version for any $\epsilon\in[0,1)$. In other words, feedback does not enlarge the $\epsilon$-capacity region of any single-destination DM-MMN consisting of independent DMCs. Thus, the strong converse also holds for the feedback version of this class of DM-MMNs.

\begin{Theorem} \label{thmPTPFeedbackVersion}
Let $(\mathcal{X}_\mathcal{I}, \mathcal{Y}_\mathcal{I}, q_{Y_\mathcal{I}|X_\mathcal{I}})$ be a DM-MMN consisting of independent DMCs with multicast demand $(\mathcal{S}, \{d\})$, and let $\mathcal{R}^{\prime}$ be the set defined in \eqref{RPrime}. Suppose $(\mathcal{X}_\mathcal{I}, \tilde {\mathcal{Y}}_\mathcal{I}, \tilde q_{\tilde Y_\mathcal{I}|X_\mathcal{I}})$ is a feedback version of $(\mathcal{X}_\mathcal{I}, \mathcal{Y}_\mathcal{I}, q_{Y_\mathcal{I}|X_\mathcal{I}})$. Let $\epsilon\in[0,1)$ be a real number and let $\mathcal{C}_\epsilon$ and $\tilde {\mathcal{C}}_\epsilon$ be the $\epsilon$-capacity regions of $(\mathcal{X}_\mathcal{I}, \mathcal{Y}_\mathcal{I}, q_{Y_\mathcal{I}|X_\mathcal{I}})$ and $(\mathcal{X}_\mathcal{I}, \tilde {\mathcal{Y}}_\mathcal{I}, \tilde q_{\tilde Y_\mathcal{I}|X_\mathcal{I}})$ respectively. Then, $\tilde {\mathcal{C}}_\epsilon =\mathcal{C}_\epsilon=\mathcal{R}^{\prime}$.
\end{Theorem}

Theorem~\ref{thmPTPFeedbackVersion} can be proved similarly to Theorem~\ref{theoremPTPNetwork}. We provide a concise proof in Appendix~\ref{appendixE}. Since the $\epsilon$-capacity region with \textit{imperfect feedback} compared with perfect feedback cannot be larger and the $\epsilon$-capacity region with \textit{no feedback} is equal to $\mathcal{R}^{\prime}$ by Theorem~\ref{theoremPTPNetwork}, it follows from Theorem~\ref{thmPTPFeedbackVersion} that the strong converse also holds for any single-destination DM-MMN consisting of independent DMCs with imperfect feedback.

\section{Conclusion and Future Work} \label{sec:conclu}
In this paper, we proved that the strong converse holds for some classes of DM-MMNs for which the cut-set bound is achievable by leveraging some elementary   properties of the conditional R\'enyi divergence. We suggest three promising avenues for future research. First, the
foremost item is to show that all rate tuples that lie in the exterior of the usual cut-set bound for DM-MMNs~\cite{elgamal_81} result in error probabilities tending to one. This seems rather challenging as we have to assert the existence of a {\em common} distribution $\bar{p}_{X_{\cal I}, Y_{\cal I}}$ for  all cut-sets $T$ in \eqref{thmMainResultfinalStatement2}. This would allow us to swap the intersection and union in Theorem \ref{thmMainResult}. Second, and less ambitiously, we also hope to extend our result to Gaussian networks \cite[Chapter 19]{elgamal}, which may be tractable if we  restrict the models under consideration to the class of Gaussian networks for which the optimum input distribution is a multivariate Gaussian. Finally, it may be fruitful and instructive to focus our attention on smaller DM-MMNs such as the DM-RC.

\appendices
\section{Proof of Proposition~\ref{propositionDlambdaToMutualInfo}} \label{appendixA}
\begin{IEEEproof}[Proof of Proposition~\ref{propositionDlambdaToMutualInfo}]
For any random variables $U$ and $V$, we let $$\mathcal{S}_{U|v} \triangleq \{u\in\mathcal{U}: \Pr\{U=u|V=v\}>0\}$$ be the {\em support} of $U$ conditioned on the event $\{V=v\}$. If $V$ is a trivial random variable, i.e., $\mathcal{V}=\emptyset$, then $\mathcal{S}_U$ is simply the support of $U$.  If $\lambda=1$, the statement of the proposition is obvious so henceforth, we prove the statement for  $\lambda\in (1,5/4]$.  Suppose $(X,Y,Z)$ is jointly distributed according to $p_{X,Y,Z}(x,y,z)$ which we abbreviate as $p(x,y,z)$ in this proof. Let \[
g(\lambda)\triangleq \log\sum_{z\in \mathcal{S}_Z}p(z)\sum_{(x,y)\in \mathcal{S}_{X,Y|z}}\frac{(p(x,y|z))^\lambda}{(p(x|z)p(y|z))^{\lambda-1}}
 \]
 be a function of $\lambda$ defined on $[1,\infty)$. Straightforward calculations involving l'H\^{o}pital's rule reveal that $g(1)=0$ and $g^\prime(1)= D(p_{X,Y|Z}||p_{X|Z}p_{Y|Z}|p_Z)$ (cf.\ Definition~\ref{defRenyiDivergence}). Using Taylor's theorem, we obtain
$$g(\lambda) = g(1) + (\lambda-1)g^\prime(1) + (\lambda-1)^2 \frac{g^{\prime\prime}(a)}{2}$$ for some $a\in [1, \lambda]$, which implies that
 \begin{equation}
g(\lambda) = (\lambda-1)D(p_{X,Y|Z}||p_{X|Z}p_{Y|Z}|p_Z) + (\lambda-1)^2 \frac{g^{\prime\prime}(a)}{2}. \label{propositionDlambdaToMutualInfoEq1}
\end{equation}
Using standard calculus techniques, we obtain
\begin{align}
g^{\prime\prime}(a)& = \frac{\sum\limits_{z\in \mathcal{S}_Z}p(z)\sum\limits_{(x,y)\in \mathcal{S}_{X,Y|z}}\frac{(p(x,y|z))^a}{(p(x|z)p(y|z))^{a-1}}\left(\log\frac{p(x,y|z)}{p(x|z)p(y|z)}\right)^2}{\sum\limits_{z\in \mathcal{S}_Z}p(z)\sum\limits_{(x,y)\in \mathcal{S}_{X,Y|z}}\frac{(p(x,y|z))^a}{(p(x|z)p(y|z))^{a-1}}} \notag\\
 & \qquad -\left(\frac{\sum\limits_{z\in \mathcal{S}_Z}p(z)\sum\limits_{(x,y)\in \mathcal{S}_{X,Y|z}}\frac{(p(x,y|z))^a}{(p(x|z)p(y|z))^{a-1}}\log\frac{p(x,y|z)}{p(x|z)p(y|z)}}{\sum\limits_{z\in \mathcal{S}_Z}p(z)\sum\limits_{(x,y)\in \mathcal{S}_{X,Y|z}}\frac{(p(x,y|z))^a}{(p(x|z)p(y|z))^{a-1}}}\right)^2. \label{propositionDlambdaToMutualInfoEq2}
\end{align}
In order to obtain an upper bound for $|g^{\prime\prime}(a)|$, we will calculate a lower bound for
\[\sum_{(x,y)\in \mathcal{S}_{X,Y|z}}\frac{(p(x,y|z))^a}{(p(x|z)p(y|z))^{a-1}}\] and upper bounds for
\[\sum_{(x,y)\in \mathcal{S}_{X,Y|z}}\frac{(p(x,y|z))^a}{(p(x|z)p(y|z))^{a-1}}\log \frac{p(x,y|z)}{p(x|z)p(y|z)}\] and \[\sum_{(x,y)\in \mathcal{S}_{X,Y|z}}\frac{(p(x,y|z))^a}{(p(x|z)p(y|z))^{a-1}}\left(\log \frac{p(x,y|z)}{p(x|z)p(y|z)}\right)^2.\] Consider the following chain of inequalities:
\begin{align}
\sum_{(x,y)\in \mathcal{S}_{X,Y|z}}\frac{(p(x,y|z))^a}{(p(x|z)p(y|z))^{a-1}} &\ge \sum_{(x,y)\in \mathcal{S}_{X,Y|z}}(p(x,y|z))^a \notag\\
& \ge \max_{x,y}(p(x,y|z))^a \notag\\
& \ge (|\mathcal{X}||\mathcal{Y}|)^{-a} \notag\\
& \ge (|\mathcal{X}||\mathcal{Y}|)^{-5/4}. \label{propositionDlambdaToMutualInfoEq3}
\end{align}
On the other hand, fix $x$, $y$ and $z$ such that $p(z)>0$ and $p(x,y|z)>0$, and consider $\frac{(p(x,y|z))^a}{(p(x|z)p(y|z))^{a-1}}\log \frac{p(x,y|z)}{p(x|z)p(y|z)}$ as well as $\frac{(p(x,y|z))^a}{(p(x|z)p(y|z))^{a-1}}\left(\log \frac{p(x,y|z)}{p(x|z)p(y|z)}\right)^2$. Since $\min\{p(x|z), p(y|z)\}\ge p(x,y|z)$, there exist $0\le k_1\le 1$ and $0\le k_2\le 1$ such that $p(x|z)=(p(x,y|z))^{k_1}$ and $p(y|z)=(p(x,y|z))^{k_2}$. Using the facts that $a\in (1,5/4]$ and $0\le k_1+k_2 \le 2$, we have
\begin{equation}
|1-(k_1+k_2)| \le 1 \label{propositionDlambdaToMutualInfoTemp1}
\end{equation}
and
\begin{equation}
a-(a-1)(k_1+k_2) \ge 1/2 \label{propositionDlambdaToMutualInfoTemp2}
\end{equation}
 Then,
\begin{align}
\frac{(p(x,y|z))^a}{(p(x|z)p(y|z))^{a-1}}\log \frac{p(x,y|z)}{p(x|z)p(y|z)} & = (p(x,y|z))^{a-(a-1)(k_1+k_2)}\log  (p(x,y|z))^{1-(k_1+k_2)}\notag\\
& \stackrel{\text{(a)}}{\le}  (p(x,y|z))^{1/2}\log  (p(x,y|z))^{-1} \notag\\
& = 2(p(x,y|z))^{1/2}\log  (p(x,y|z))^{-1/2} \notag\\
& \stackrel{\text{(b)}}{\le} 2 e^{-1}\log  e \label{propositionDlambdaToMutualInfoTemp3}
\end{align}
and
\begin{align}
\frac{(p(x,y|z))^a}{(p(x|z)p(y|z))^{a-1}}\left(\log \frac{p(x,y|z)}{p(x|z)p(y|z)}\right)^2 & = (p(x,y|z))^{a-(a-1)(k_1+k_2)}(\log  (p(x,y|z))^{1-(k_1+k_2)})^2\notag\\
& \stackrel{\text{(c)}}{\le}  (p(x,y|z))^{1/2}(\log  (p(x,y|z))^{-1})^2 \notag\\
& = 4(p(x,y|z))^{1/2}(\log  (p(x,y|z))^{-1/2})^2 \notag\\
& \stackrel{\text{(d)}}{\le} 16 e^{-2}\log  e, \label{propositionDlambdaToMutualInfoTemp4}
\end{align}
where
\begin{enumerate}
\item[(a)] follows from \eqref{propositionDlambdaToMutualInfoTemp1} and \eqref{propositionDlambdaToMutualInfoTemp2};
    \item[(b)] follows from calculus that $q\log  q^{-1} \le e^{-1}\log  e$ for all $0< q\le 1$;
    \item[(c)] follows from \eqref{propositionDlambdaToMutualInfoTemp1} and \eqref{propositionDlambdaToMutualInfoTemp2};
    \item[(d)] follows from calculus that $q(\log  q^{-1})^2 \le 4e^{-2}\log  e$ for all $0< q\le 1$.
\end{enumerate}
Combining \eqref{propositionDlambdaToMutualInfoEq1}, \eqref{propositionDlambdaToMutualInfoEq2}, \eqref{propositionDlambdaToMutualInfoEq3}, \eqref{propositionDlambdaToMutualInfoTemp3} and  \eqref{propositionDlambdaToMutualInfoTemp4}, we obtain
\begin{align*}
g(\lambda) \le (\lambda-1)D(p_{X,Y|Z}||p_{X|Z}p_{Y|Z}|p_Z) + 8(\lambda-1)^2 (|\mathcal{X}||\mathcal{Y}|)^{5},
\end{align*}
which implies that for each $\lambda\in (1, 5/4]$ (note $\lambda\ne 1$ so we can cancel the common factors $\lambda-1$),
\[
D_\lambda(p_{X,Y|Z}||p_{X|Z}p_{Y|Z}|p_Z) \le D(p_{X,Y|Z}||p_{X|Z}p_{Y|Z}|p_Z) +8(\lambda-1) (|\mathcal{X}||\mathcal{Y}|)^{5}
\]
and hence \eqref{propositionDlambdaToMutualInfoStatement} follows.
\end{IEEEproof}
\section{Proof of Corollary~\ref{corollaryProductDistributionOuterBound}} \label{appendixB}
\begin{IEEEproof}[Proof of Corollary~\ref{corollaryProductDistributionOuterBound}]
Suppose the DM-MMN is dominated by some maximal product distribution $p_{X_\mathcal{I}}^*\triangleq \prod_{i=1}^N p_{X_i}^*$ such that for each $T\subseteq \mathcal{I}$, we have
\begin{align*}
 I_{p_{X_{\mathcal{I}}}^*q_{Y_{T^c}|X_\mathcal{I}}}(X_T; Y_{T^c}|X_{T^c})=  \max_{p_{X_\mathcal{I}}} \{ I_{p_{X_{\mathcal{I}}}q_{Y_{T^c}|X_\mathcal{I}}}(X_T; Y_{T^c}|X_{T^c})\}.
\end{align*}
This then implies from Theorem~\ref{thmMainResult} that for each $\epsilon \in [0,1)$,
\begin{align*}
\mathcal{C}_\epsilon &\subseteq \bigcap_{T\subseteq \mathcal{I}: T^c \cap \mathcal{D} \ne \emptyset }  \left\{ R_\mathcal{I}\left| \: \parbox[c]{2.4in}{$ \sum_{ i\in T} R_{i}
 \le  I_{p_{X_{\mathcal{I}}}^*q_{Y_{T^c}|X_\mathcal{I}}}(X_T; Y_{T^c}|X_{T^c}),\\
 R_i=0 \text{ for all }i\in\mathcal{S}^c$} \right.\right\} \stackrel{\eqref{RoutDet}}{\subseteq}\mathcal{R}_{\text{out}}^* .
\end{align*}
This completes the proof.
\end{IEEEproof}
\section{Proof of Corollary~\ref{corollaryInnerBoundPTP}}\label{appendixC}
\begin{IEEEproof}[Proof of Corollary~\ref{corollaryInnerBoundPTP}]
Construct a counterpart of the channel $(\mathcal{X}_\mathcal{I}, \mathcal{Y}_\mathcal{I}, q_{Y_\mathcal{I}|X_\mathcal{I}})$ as follows: Let $(\bar{\mathcal{X}}_\mathcal{I}, \bar{\mathcal{X}}_\mathcal{I}, \bar q_{\bar X_\mathcal{I}|\bar X_\mathcal{I}})$ be a noiseless DM-MMN consisting of independent DMCs with multicast demand $(\mathcal{S}, \mathcal{D})$ such that for each $(i,j)\in \mathcal{I}\times \mathcal{I}$, the DMC carrying information from node~$i$ to node~$j$ is an error-free (noiseless) channel, denoted by $\bar q_{\bar X_{i,j} | \bar X_{i,j}}$, with capacity $C_{i,j}$ (cf.\ \eqref{defPTPCapacity}). To be more precise, $\bar q_{\bar X_{i,j} | \bar X_{i,j}}$ can carry $\lfloor n C_{i,j} \rfloor$ error-free bits for each $(i,j)\in \mathcal{I}\times \mathcal{I}$ for $n$ uses of $(\bar{\mathcal{X}}_\mathcal{I}, \bar{\mathcal{X}}_\mathcal{I}, \bar q_{\bar X_\mathcal{I}|\bar X_\mathcal{I}})$. Let $\bar {\mathcal{C}}$ denote the capacity region of $(\bar{\mathcal{X}}_\mathcal{I}, \bar{\mathcal{X}}_\mathcal{I}, \bar q_{\bar X_\mathcal{I}|\bar X_\mathcal{I}})$. It follows from the network equivalence theory \cite{networkEquivalencePartI} that $\mathcal{C}_0 = \bar{\mathcal{C}}$. In addition, it has been shown in \cite[Section IIA]{noisyNetworkCoding} that $\bar{\mathcal{C}}=\mathcal{R}^{\prime}$.
 Consequently, $\mathcal{R}^{\prime}=\bar{\mathcal{C}} = \mathcal{C}_0$, which is what was to be proved.
\end{IEEEproof}
\section{Proof of Lemma~\ref{lemmaPTP}} \label{appendixD}
\begin{IEEEproof}[Proof of Lemma~\ref{lemmaPTP}]
Since $\mathcal{C}_\epsilon \subseteq \mathcal{R}_{\text{out}}$ for all $\epsilon \in [0,1)$ by Theorem~\ref{thmMainResult}, it remains to show $\mathcal{R}_{\text{out}} \subseteq \mathcal{R}^{\prime}$. In order to obtain an outer bound of $\mathcal{R}_{\text{out}}$, we consider the following chain of inequalities for each $p_{X_\mathcal{I}}$ and each $T\subseteq \mathcal{I}$: 
 \begin{align}
& I_{p_{X_{\mathcal{I}}}q_{Y_{T^c}|X_\mathcal{I}}}(X_T; Y_{T^c}|X_{T^c}) \notag\\
 & = \sum_{j\in T^c} I_{p_{X_{\mathcal{I}}}q_{Y_{T^c}|X_\mathcal{I}}}(X_T; Y_{j}|X_{T^c}, Y_{\{\bar j \in T^c: \bar j<j\}}) \notag\\
 &\stackrel{\eqref{alphabetYSequence}}{=} \sum_{j\in T^c} \sum_{\ell=1}^N I_{p_{X_{\mathcal{I}}}q_{Y_{T^c}|X_\mathcal{I}}}(X_T; Y_{\ell, j}|X_{T^c}, Y_{\{\bar j \in T^c: \bar j<j\}}, \{Y_{m, j}\}_{m=1}^{\ell -1}) \notag\\
  &\stackrel{\text{(a)}}{=} \sum_{j\in T^c} \sum_{\ell\in T} I_{p_{X_{\mathcal{I}}}q_{Y_{T^c}|X_\mathcal{I}}}(X_T; Y_{\ell, j}|X_{T^c}, Y_{\{\bar j \in T^c: \bar j<j\}}, \{Y_{m, j}\}_{m=1}^{\ell -1})\notag\\
   &\stackrel{\text{(b)}}{=} \sum_{j\in T^c} \sum_{\ell\in T} I_{p_{X_{\mathcal{I}}}q_{Y_{T^c}|X_\mathcal{I}}}(X_{\ell,j}; Y_{\ell, j})\notag\\
   &\stackrel{\eqref{defPTPCapacity}}{\le} \sum_{(\ell,j)\in T\times T^c} C_{\ell,j}. \label{PTPinLemmaEq1}
 \end{align}
 where
 \begin{enumerate}
 \item[(a)] follows from the fact that for all $\ell \in T^c$, \begin{align*}
    & I_{p_{X_{\mathcal{I}}}q_{Y_{T^c}|X_\mathcal{I}}}(X_T; Y_{\ell, j}|X_{T^c}, Y_{\{\bar j \in T^c: \bar j<j\}}, \{Y_{m, j}\}_{m=1}^{\ell -1}) \notag\\
    & = H_{p_{X_{\mathcal{I}}}q_{Y_{T^c}|X_\mathcal{I}}}(Y_{\ell, j}|X_{T^c}, Y_{\{\bar j \in T^c: \bar j<j\}}, \{Y_{m, j}\}_{m=1}^{\ell -1})-H_{p_{X_{\mathcal{I}}}q_{Y_{T^c}|X_\mathcal{I}}}(Y_{\ell, j}|X_{\mathcal{I}}, Y_{\{\bar j \in T^c: \bar j<j\}}, \{Y_{m, j}\}_{m=1}^{\ell -1}) \notag\\
    & \stackrel{\eqref{markovChainPTPNetwork}}{=} H_{p_{X_{\mathcal{I}}}q_{Y_{T^c}|X_\mathcal{I}}}(Y_{\ell, j}|X_{\ell, j})- H_{p_{X_{\mathcal{I}}}q_{Y_{T^c}|X_\mathcal{I}}}(Y_{\ell, j}|X_{\ell, j}) \notag\\
    &=0;
     \end{align*}
      \item[(b)] follows from the fact that for all $\ell \in T$,
      \begin{align*}
    & I_{p_{X_{\mathcal{I}}}q_{Y_{T^c}|X_\mathcal{I}}}(X_T; Y_{\ell, j}|X_{T^c}, Y_{\{\bar j \in T^c: \bar j<j\}}, \{Y_{m, j}\}_{m=1}^{\ell -1}) \notag\\
    & \le H_{p_{X_{\mathcal{I}}}q_{Y_{T^c}|X_\mathcal{I}}}(Y_{\ell, j})-H_{p_{X_{\mathcal{I}}}q_{Y_{T^c}|X_\mathcal{I}}}(Y_{\ell, j}|X_{\mathcal{I}}, Y_{\{\bar j \in T^c: \bar j<j\}}, \{Y_{m, j}\}_{m=1}^{\ell -1}) \notag\\
    & \stackrel{\eqref{markovChainPTPNetwork}}{=} H_{p_{X_{\mathcal{I}}}q_{Y_{T^c}|X_\mathcal{I}}}(Y_{\ell, j})- H_{p_{X_{\mathcal{I}}}q_{Y_{T^c}|X_\mathcal{I}}}(Y_{\ell, j}|X_{\ell, j}) \notag\\
    &=I(X_{\ell,j}; Y_{\ell,j}).
     \end{align*}
 \end{enumerate}
 Combining \eqref{Rout}, \eqref{RPrime} and \eqref{PTPinLemmaEq1}, we have $ \mathcal{R}_{\text{out}}\subseteq \mathcal{R}^\prime$. \end{IEEEproof}

\section{Proof of Theorem~\ref{thmPTPFeedbackVersion}} \label{appendixE}
\begin{IEEEproof}[Proof of Theorem~\ref{thmPTPFeedbackVersion}]
Fix any $\epsilon \in [0,1)$. Since $\mathcal{C}_\epsilon=\mathcal{R}^{\prime}$ by Theorem~\ref{theoremPTPNetwork} and $\mathcal{C}_\epsilon \subseteq \tilde {\mathcal{C}}_\epsilon$, it remains to show that $\tilde {\mathcal{C}}_\epsilon \subseteq \mathcal{R}^{\prime}$. Define
 \begin{equation}
 \tilde{\mathcal{R}}_{\text{out}} \triangleq  \bigcap_{T\subseteq \mathcal{I}: T^c \cap \{d\} \ne \emptyset } \bigcup_{ p_{X_{\mathcal{I}}}} \left\{ R_\mathcal{I}\left| \: \parbox[c]{2.4in}{$ \sum_{ i\in T} R_{i}
 \le  I_{p_{X_{\mathcal{I}}}\tilde q_{\tilde Y_{T^c}|X_\mathcal{I}}}(X_T; \tilde Y_{T^c}|X_{T^c}),\\
 R_i=0 \text{ for all }i\in\mathcal{S}^c$} \right.\right\}. \label{RoutTilde}
 \end{equation}
 Since $\tilde{\mathcal{C}}_\epsilon \subseteq \tilde{\mathcal{R}}_{\text{out}}$ for all $\epsilon \in [0,1)$ by Theorem~\ref{thmMainResult} and $\mathcal{R}_{\text{out}} \subseteq \mathcal{R}^{\prime}$ by Lemma~\ref{lemmaPTP}, it suffices to show $\tilde{\mathcal{R}}_{\text{out}} = \mathcal{R}_{\text{out}} $. To this end, we consider the following chain of equalities for each $p_{X_\mathcal{I}}$ and each $T\subseteq \mathcal{I}$ such that $T^c \cap \{d\}\ne \emptyset$:
 \begin{align}
 I_{p_{X_{\mathcal{I}}}\tilde q_{\tilde Y_{T^c}|X_\mathcal{I}}}(X_T; \tilde Y_{T^c}|X_{T^c}) &\stackrel{\text{(a)}}{=} I_{p_{X_{\mathcal{I}}}q_{Y_{T^c}|X_\mathcal{I}}}(X_T; Y_{T^c}, Y_d|X_{T^c}) \notag\\
 & \stackrel{\text{(b)}}{=}I_{p_{X_{\mathcal{I}}}q_{Y_{T^c}|X_\mathcal{I}}}(X_T; Y_{T^c}|X_{T^c}). \label{PTPinLemmaEq1**}
 \end{align}
 where
 \begin{enumerate}
 \item[(a)] follows from Condition 2 in Definition~\ref{defDM-MMNwithFeedback};
 \item[(b)] follows from the fact that $T^c \cap \{d\}\ne \emptyset$.
 \end{enumerate}
 Combining \eqref{Rout}, \eqref{RoutTilde} and \eqref{PTPinLemmaEq1**}, we have $ \mathcal{R}_{\text{out}}= \tilde{ \mathcal{R}}_{\text{out}}$.
 \end{IEEEproof}

\ifCLASSOPTIONcaptionsoff
  \newpage
\fi



%
%
%
\section*{Acknowledgments}
The authors are indebted to Prof.\ Shun Watanabe for pointing out an error in an earlier version of this paper.


%
%
%
%
%




\end{document}